\documentclass[twocolumn]{aastex631}

\usepackage{amsmath}
\usepackage{mathtools}
\usepackage{multirow}
\usepackage{tabularx}
\usepackage{natbib}
\defcitealias{2019MNRAS.485..248G}{G19}

\begin{document}

\title{Identifying Host Galaxies of Supermassive Black Hole Binaries Found by PTAs}

\author[0000-0001-5681-4319]{Polina Petrov}
\affiliation{Department of Physics and Astronomy, Vanderbilt University, 2301 Vanderbilt Place, Nashville, TN 37235, USA}

\author[0000-0001-8217-1599]{Stephen R. Taylor}
\affiliation{Department of Physics and Astronomy, Vanderbilt University, 2301 Vanderbilt Place, Nashville, TN 37235, USA}

\author[0000-0003-3579-2522]{Maria Charisi}
\affiliation{Department of Physics and Astronomy, Vanderbilt University, 2301 Vanderbilt Place, Nashville, TN 37235, USA}
\affiliation{Department of Physics and Astronomy, Washington State University, Pullman, WA 99163, USA}
\affiliation{Institute of Astrophysics, FORTH, GR-71110, Heraklion, Greece}

\author[0000-0002-4430-102X]{Chung-Pei Ma}
\affiliation{Department of Astronomy, University of California, Berkeley, 501 Campbell Hall 3411, Berkeley, CA 94720, USA}
\affiliation{Department of Physics, University of California, Berkeley, CA 94720, USA}

\begin{abstract}

Supermassive black hole binaries (SMBHBs) are thought to form in galaxy mergers, possessing the potential to produce electromagnetic (EM) radiation as well as gravitational waves (GWs) detectable with pulsar timing arrays (PTAs). Once GWs from individually resolved SMBHBs are detected, the identification of the host galaxy will be a major challenge due to the ambiguity in possible EM signatures and the poor localization capability of PTAs. To aid EM observations in choosing follow-up sources, we use NANOGrav's galaxy catalog to quantify the number of plausible hosts in both realistic and idealistic scenarios. We outline a host identification pipeline that injects a single-source GW signal into a simulated PTA dataset, recovers the signal using production-level techniques, quantifies the localization region and number of galaxies contained therein, and finally imposes cuts on the galaxies using parameter estimates from the GW search. In an ideal case, the 90\% credible areas span 29 deg$^2$ to 241 deg$^2$, containing about 14 to 341 galaxies. After cuts, the number of galaxies remaining ranges from 22 at worst to 1 true host at best. In a realistic case, these areas range from 287 deg$^2$ to 530 deg$^2$ and enclose about 285 to 1238 galaxies. After cuts, the number of galaxies is 397 at worst and 27 at best. While the signal-to-noise ratio is the primary determinant of the localization area of a given source, we find that the area is also influenced by the proximity to nearby pulsars on the sky and the binary chirp mass.

\end{abstract}


\section{Introduction}
\label{sec:intro}

Supermassive black hole binaries (SMBHBs) are thought to form and become gravitationally bound following the merger of two galaxies \citep{1980Natur.287..307B}. The binary orbit evolves through physical mechanisms such as dynamical friction, stellar scatterings, and torques from a circumbinary disk, until the black holes are sufficiently close that gravitational wave (GW) emission becomes an efficient method for radiating away energy. SMBHBs with masses of $10^8$ – $10^{10} M_{\odot}$ and sub-parsec separations should emit GWs at nanohertz frequencies $\sim10^{-9}$ – $10^{-7}$ Hz that can be detected with pulsar timing arrays (PTAs).

PTAs aim to detect GWs using dozens of extremely well-timed pulsars to look for correlated deviations in the arrival times of radio pulses \citep{1978SvA....22...36S, 1979ApJ...234.1100D, 1990ApJ...361..300F}. PTAs around the world, including the North American Nanohertz Observatory for Gravitational Waves (NANOGrav), the European Pulsar Timing Array (EPTA), the Indian Pulsar Timing Array (InPTA), the Parkes Pulsar Timing Array (PPTA) and the Chinese Pulsar Timing Array (CPTA), have recently found varying levels of evidence for a stochastic gravitational wave background (GWB) \citep{2023ApJ...951L...8A, 2023A&A...678A..50E, 2023ApJ...951L...6R, CPTA_GWB}. These datasets will be combined with data from the MeerKAT Pulsar Timing Array (MPTA; \citealt{2023MNRAS.519.3976M}) and reanalyzed as part of the third data release from the International PTA (IPTA-DR3), which is expected to provide more conclusive evidence than any individual PTA's dataset \citep{2023arXiv230900693T}.

The dominant source of the GWB is expected to be the superposition of millions of GW signals emanating from a cosmic population of SMBHBs. PTAs should also detect GWs emitted by individual binaries in the local Universe that are sufficiently ``loud” to be resolvable above the GWB \citep{2009MNRAS.394.2255S}. Called continuous waves (CWs) due to their minimal frequency evolution, these GW sources may be detected within the decade \citep{2015MNRAS.451.2417R, 2017NatAs...1..886M, 2018MNRAS.477..964K, 2022ApJ...941..119B}. Such a detection of an individually resolvable binary will have significant impacts on GW astrophysics in that it will not only provide us with compelling evidence that nanohertz GWs indeed come from SMBHBs, but it will also kick off the pursuit of a multi-messenger system.

Because SMBHBs are thought to form as a natural consequence of galaxy mergers, many of these systems may be embedded in gaseous environments \citep{1992ARA&A..30..705B,2005ApJ...620L..79S,2006ApJ...645..986R} and have the potential to produce electromagnetic (EM) radiation. However, several questions surrounding the EM counterparts of SMBHBs remain unanswered. Do all SMBHBs produce light that is easily observable? Is the EM signature of such binaries unique? We may find SMBHBs in a variety of systems, and each type of system may generate a distinct EM signature, or none at all. A wide range of possible EM signatures have been proposed, including doppler-shifted broad emission lines in quasar spectra \citep{1968SvA....12..401K,1980Natur.287..307B}, periodic brightness variations in quasar light curves \citep{2009ApJ...700.1952H}, imprints in the morphology of radio jets \citep{1980Natur.287..307B}, deficits in spectral energy distributions \citep{2012ApJ...761...90G, 2014ApJ...785..115R}, and changes in the spectral line profile of Fe K-$\alpha$ \citep{2013MNRAS.432.1468M}, among others \citep[for a review see, e.g.,][]{2022LRR....25....3B,2023arXiv231016896D}. Many of these signatures can also be explained by other physical processes or may be produced in single-SMBH systems.

Given that we expect SMBHBs to reside in massive post-merger galaxies, some studies have focused efforts on identifying the galaxy hosting the binary \citep{2014ApJ...784...60S, 2014MNRAS.439.3986R, 2019MNRAS.485..248G}. Nonetheless, host galaxy identification continues to be a challenge due to the ambiguity surrounding EM signatures as well as the sky localization that PTAs can achieve. The first observations of individual binaries are predicted to have very poor localization, with sky areas spanning $\sim$ $10^2$ – $10^3$ deg$^2$ \citep{2010PhRvD..81j4008S, 2016ApJ...817...70T, 2019MNRAS.485..248G}. Areas of this size will contain thousands of plausible host galaxies, making EM follow-up observations on every galaxy impractical in terms of telescope time and resources.

GW signal localization is not a unique problem to PTA astrophysics, but is rather a broad issue that affects detectors across the GW spectrum. Ground-based detectors LIGO, Virgo, and KAGRA (LVK) have similarly contended with large localization regions, and the future space-based Laser Interferometer Space Antenna (LISA) will face the same problem \citep{2020PhRvD.102h4056M, 2023MNRAS.519.5962L}. For sources seen by LVK, a number of catalogs have been assembled to aid in the search for host galaxies and potential EM counterparts \citep[e.g., GLADE+;][] {2022MNRAS.514.1403D}, extending out to a distance of $\sim$ 100 Mpc and concentrating more specifically on star-forming galaxies. NANOGrav has similarly compiled a catalog of massive galaxies in the local universe \citep[][discussed in more detail in Section \ref{sec:gxycat}]{2021ApJ...914..121A}. These studies and tools have been essential steps towards connecting GW signals to their host galaxies. Continued development of host galaxy identification methods will be crucial to achieving a coordinated multi-messenger detection of a SMBHB system, especially as PTAs become increasingly more sensitive over the next decade.

In this paper, we make strides towards quantifying SMBHB host identification prospects and reducing the number of plausible hosts to a more manageable size for EM follow-up and multi-messenger studies. We outline a pipeline that mimics the discovery process for an individually resolvable binary, simulating an IPTA-style array, injecting putative binaries in various host galaxies, searching for and recovering the GW signals, quantifying the localization areas and potential hosts therein, and finally implementing cuts on those hosts based on the posterior distributions of the recovered signals.

This paper is organized as follows: in Section \ref{sec:methods} we discuss the signal model, the galaxy catalog containing our potential hosts, the nature of our simulated datasets, and the host identification pipeline. In Section \ref{sec:results} we present our results on nine fiducial injections as well as related case studies. In Section \ref{sec:discuss} we discuss the implications of our results and plans for future work, and in Section \ref{sec:concl} we summarize our conclusions. In all of the following we assume natural units with $G=c=1$, and all GW equations assume General Relativity.

\section{Methods} \label{sec:methods}

Here we detail each of the components involved in our signal simulation, discovery, and host reduction pipeline. In Section \ref{sec:model} we start with the mathematical formalism for GWs emitted by a SMBHB, followed by the GW signal search and analysis. Section \ref{sec:gxycat} includes a description of the galaxy catalog employed in this work. Section \ref{sec:sims} covers our simulation setup, including the array configuration, the host galaxies we choose to inject with a CW signal, and the injected parameters. Finally, in Section \ref{sec:cuts} we outline the ways in which we impose cuts on potential host galaxies.

\subsection{Gravitational Waves from a Supermassive Black Hole Binary} \label{sec:model}

\subsubsection{Signal Model}\label{sec:model1}

PTA observations are made in the form of pulse times-of-arrival (TOAs). From each pulsar's measured TOAs, we subtract the pulsar's best-fit timing model, which includes any deterministic factors that can influence the TOAs, such as the pulsar spin period, proper motion, orbital parameters for binary pulsars, etc. The deviations resulting from this subtraction are known as the timing residuals, and should be produced only by GWs and any sources of noise. We can then describe the influence of GWs on the pulsar's TOAs beginning with the residuals, which are modeled in each pulsar as the vector

\begin{equation}
    \delta\boldsymbol{t} =  \mathit{M}\boldsymbol{\epsilon} + \boldsymbol{\mathit{n}}_\mathrm{WN} + \boldsymbol{\mathit{n}}_\mathrm{RN} + \boldsymbol{\mathit{n}}_\mathrm{GWB} + \boldsymbol{\mathit{s}}.
\end{equation}

\noindent The matrix $M$ contains partial derivatives of the pulsar's TOAs with respect to each timing model parameter, and $\boldsymbol{\epsilon}$ is a vector of linearized timing model parameter offsets from the fitting solution. 

The vector $\boldsymbol{n}_{\mathrm{WN}}$ describes the pulsar's white noise and consists of an extra correction factor (EFAC), a multiplicative factor that adjusts the TOA uncertainties. The vector $\boldsymbol{n}_{\mathrm{RN}}$ describes the pulsar's intrinsic red noise, whose power spectral density is modeled as

\begin{equation}
P = \frac{A_{\mathrm{RN}}^2}{12\pi^2}\left(\frac{f}{f_{\mathrm{yr}}}\right)^{-\gamma_{\mathrm{RN}}} \mathrm{yr}^3,
\end{equation}

\noindent where $A_{\mathrm{RN}}$ is the red noise amplitude, $f_{\mathrm{yr}}$ is 1/(1yr) in Hz, and $\gamma_{\mathrm{RN}}$ is the power law spectral index. The vector $\boldsymbol{n}_{\mathrm{GWB}}$ represents the GWB signal present in every pulsar---albeit modulated by directional response factors---and, similar to the pulsar red noise, has a power-spectral density modeled as a power-law red noise process of the form

\begin{equation}
P = \frac{A_{\mathrm{GWB}}^2}{12\pi^2}\left(\frac{f}{f_{\mathrm{yr}}}\right)^{-\gamma_{\mathrm{GWB}}} \mathrm{yr}^3,
\end{equation}

\noindent where the amplitude $A_{\mathrm{GWB}}$ and the spectral index $\gamma_{\mathrm{GWB}}$ are common to all of the pulsars in the array. Although the GWB will induce correlated timing offsets between pulsars, we do not include this effect as part of our model for either injections or recoveries (the reasoning for which we discuss in the following section). 

Finally, the timing deviation signal caused by an individual binary is represented by the vector $\boldsymbol{s}$, which can be written as

\begin{equation} \label{eq:s_t}
\begin{split}
    s(t) = F^+(\theta, \phi, \psi)[s_+(t_p) - s_+(t)] \\
    + F^{\times}(\theta, \phi, \psi)[s_{\times}(t_p) - s_{\times}(t)],
\end{split}
\end{equation}

\noindent where the antenna pattern functions $F^{+,\times}$ describe the pulsars' response to the GW source for each of the $+$ (``plus") and $\times$ (``cross") polarization modes. This function depends on the GW polarization angle $\psi$, as well as the sky location of the binary in spherical polar coordinates ($\theta$, $\phi$) \citep[for more details, see][]{2023ApJ...951L..28A}. These coordinates are related to the equatorial coordinates by ($\theta$, $\phi$) $=$ ($\pi/2 - \delta$, $\alpha$), where $\alpha$ is the right ascension and $\delta$ the declination.

The signal induced at the Earth (the ``Earth term") is denoted by $s_{+,\times}(t)$, and the signal induced at the pulsar (the ``pulsar term") is denoted by $s_{+,\times}(t_p)$, where $t$ and $t_p$ represent the time at which the GW passes the Earth (more specifically, the Solar System Barycenter) and the pulsar, respectively. These times are related to each other by

\begin{equation} \label{eq:t_p}
    t_p = t - L(1 - \cos\mu),
\end{equation}

\noindent where $L$ is the distance to the pulsar and $\mu$ is the angle between the GW origin and the pulsar's position on the sky.

For a circular binary at zeroth post-Newtonian (0-PN) order, $s_{+,\times}$ is written as

\begin{equation} \label{eq:s_+}
    s_+(t) = \frac{\mathcal{M}^{5/3}}{d_L\omega(t)^{1/3}}[-\sin2\Phi(t)(1+ \cos^2\iota)],
\end{equation}
\begin{equation} \label{eq:s_x}
    s_{\times}(t) = \frac{\mathcal{M}^{5/3}}{d_L\omega(t)^{1/3}}[2\cos2\Phi(t)\cos\iota],
\end{equation}

\noindent where $d_L$ is the luminosity distance to the source, $\mathcal{M} \equiv (m_1m_2)^{3/5} / (m_1 + m_2)^{1/5}$ is a combination of the two black hole masses $m_1$ and $m_2$ known as the chirp mass, and $\iota$ is the inclination angle of the binary, defined as the angle between the line of sight and the binary's orbital angular momentum. A face-on binary corresponds to an inclination angle of $\iota=0$, while an edge-on binary corresponds to $\iota=\pi/2$.

As the binary loses energy due to the emission of GWs, the orbital frequency evolves over time as \citep{PhysRev.131.435,PhysRev.136.B1224}
\begin{equation} \label{eq:domega_dt}
\frac{\mathrm{d}\omega}{\mathrm{d}t} = \frac{96}{5}\mathcal{M}^{5/3}\omega(t)^{11/3},
\end{equation}
such that
\begin{equation} \label{eq:omega_t}
    \omega(t) = \omega_0 \left[1 - \frac{256}{5}\mathcal{M}^{5/3}\omega_0^{8/3}(t-t_0) \right]^{-3/8},
\end{equation}

\noindent where the initial orbital frequency of the Earth term is related to the GW frequency by $\omega_0 = \omega(t_0) = \pi f_{\mathrm{GW}}$. Note that $\mathcal{M}$ and $\omega$ refer to the observer-frame quantities, which are related to the rest-frame quantities by $\mathcal{M}_r = \mathcal{M}/(1 + z)$ and $\omega_r = \omega(1 + z)$. Since the current sensitivity of PTAs is limited to individual SMBHBs in the local universe, we set $1 + z \simeq 1$. The orbital phase of the binary evolves over time as

\begin{equation} \label{eq:phi_t}
    \Phi(t) = \Phi_0 + \frac{1}{32}\mathcal{M}^{-5/3}[\omega_0^{-5/3} - \omega(t)^{-5/3}],
\end{equation}

\noindent where $\Phi_0$ is the initial orbital phase. 

As PTA experiments span only a few decades of observations, the binary evolution over the observing timespan is expected to be negligible. For example, a binary with a chirp mass of $\mathcal{M}=10^9$ $\mathrm{M}_\odot$ and orbital frequency of 10 nHz would see a change in frequency on the order of $10^{-5}$ nHz, which is much smaller than the frequency resolution of PTAs. Therefore, we do not use the full evolution expressions in \autoref{eq:omega_t} and \autoref{eq:phi_t}, but rather make the assumption that each binary emits monochromatic GWs over the timing baseline of the PTA. On the other hand, the orbital evolution \textit{is} significant when considering the light travel time between the Earth and any given pulsar. Typical pulsar distances are on the order of kiloparsecs, resulting in thousands of years of evolution between the Earth term and pulsar term frequencies. The Earth term of the signal always occurs at a later time than the pulsar term, but can be evolved backwards to obtain the pulsar term dynamical state, again via \autoref{eq:domega_dt}. 

Finally, one can define a strain amplitude quantity of the signal, $h_0$, such that \citep{2009LRR....12....2S}

\begin{equation}
h(t) = F_+h_+ + F_{\times}h_{\times} = Ah_0\cos(\Phi(t) - \Phi_0),
\end{equation}
where $A = (A_+^2 + A_\times^2)^{1/2}$, $A_+=F_+(1+\cos\iota^2)/2$, $A_\times=F_\times\cos\iota$, and $h_0$ is related to the chirp mass, GW frequency, and luminosity distance by

\begin{equation} \label{eq:h}
    h_0 = \frac{2\mathcal{M}^{5/3}(\pi f_\mathrm{GW})^{2/3}}{d_L}.
\end{equation}

\noindent The model for a CW emanating from a circular SMBHB can thus be modeled as a deterministic signal, which we describe with the eight parameters $\{\theta, \phi, \mathcal{M}, f_{\rm{GW}}, \iota, \psi, \Phi_0, h_0\}$, as well as $2N$ pulsar parameters  $\{L_i, \Phi_i\}$ for $N$ pulsars in the array, corresponding to the pulsar distance and the binary orbital phase when the GW passes by the pulsar.

\subsubsection{Signal Recovery}\label{sec:model2}

To recover the CW signal, our pipeline begins with an initial pilot search using the frequentist ${F}_p$ and ${F}_e$ detection statistics derived in \cite{2012PhRvD..85d4034B} and \cite{2012ApJ...756..175E}, both of which involve maximum-likelihood based algorithms. First, we compute the ${F}_p$ statistic as a function of the GW frequency uniformly in the range $\log_{10}(f_{\mathrm{GW}}/\rm{Hz}) \in [-9,-7]$ and select the frequency that maximizes this statistic. We find that there are typically two maximum likelihood frequencies, one corresponding to the injected Earth term frequency and the other corresponding to a lower, pulsar term frequency. Because the Earth term frequency will always be higher than the pulsar term frequency, we choose the higher of the two frequencies as our global maximum estimate.

Next, we use Markov Chain Monte Carlo (MCMC) methods to sample the ${F}_e$ statistic and estimate the source's position on the sky. The GW frequency is fixed to the global maximum value determined from the ${F}_p$ statistic, while the sky location parameters are sampled using \texttt{PTMCMCSampler} \citep{justin_ellis_2017_1037579} and uniform priors in the ranges $\cos\theta \in [-1,1]$ and $\phi \in [0,2\pi]$. The global maximum sky location is then taken to be the maximum a posteriori position. We set these maximum likelihood values of $\log_{10}f_{\mathrm{GW}}$, $\cos\theta$, and $\phi$ as the initial positions in the binary parameter estimation stage to promote better subsequent MCMC sampler convergence, but we note that these parameters are indeed explored along with all others as described next.

We use the MCMC sampler \texttt{QuickCW} \citep{2022PhRvD.105l2003B}, built on top of the \texttt{enterprise} software package \citep{2019ascl.soft12015E}, to estimate the binary parameters. While \texttt{enterprise} constructs the priors and the signal model, \texttt{QuickCW}
employs a custom likelihood calculation using a Metropolis-within-Gibbs sampler and the Multiple-Try MCMC technique. As CW searches involve complex, high-dimensional parameter spaces, we additionally use parallel tempering to aid in sampling.

We sample over the $8+2N$ CW parameters, whose priors can be found in \autoref{tab:priortable}. We fix the white noise term EFAC=1, and the intrinsic red noise parameters of each pulsar are fixed at the best-fit values listed in their respective dataset papers (see Section \ref{sec:sims} for information about pulsar properties taken from the NANOGrav 15yr, PPTA DR3, EPTA+InPTA DR2new+, and MPTA DR1 dataset papers). The GWB is modeled as a common uncorrelated red noise process present across all pulsars, with a fixed amplitude $A_{\mathrm{GWB}} = 1.92 \times 10^{-15}$ and power law spectral index $\gamma_{\mathrm{GWB}} = 13/3$. These values are taken from the NANOGrav 12.5-year dataset, as this dataset serves as the basis of our simulations (discussed in Section~\ref{sec:sims}). We do not include GWB-induced spatial correlations between pulsars for two key reasons: $(i)$ this slows the analysis considerably for each run in our large suite of simulations, and the influence of GWB spatial correlations with CW recovery is not the focus of this study; $(ii)$ \texttt{QuickCW} does not yet include GWB correlations in its model (although post-processing techniques involving sample reweighting have been developed \citep{2023PhRvD.107h4045H}).

While we do not sample the luminosity distance directly, we can calculate the effective prior using \autoref{eq:h} along with our chirp mass, frequency, and strain priors. Similarly, one can plug the relevant parameters' posterior samples into a rearranged \autoref{eq:h} to obtain posteriors on the luminosity distance. We apply an astrophysically-motivated yet sufficiently uninformative log-uniform prior on the luminosity distance, $\log_{10}(d_L/\mathrm{Mpc}) \in [1,4]$, to our posteriors via reweighting; this is achieved by assigning a weight to posterior samples corresponding to the ratio of this new $\log_{10}d_L$ prior over the old prior, followed by resampling with replacement of weighted samples.

\begin{deluxetable}{cc}
\setlength{\tabcolsep}{1.5em}
\tablecaption{CW priors used for all analyses in this work.\label{tab:priortable}}
\tablecolumns{2}
\tablehead{
\colhead{Parameter} & \colhead{Prior}
}
\startdata
$\cos\theta$ & Uniform(-1, 1) \\
$\phi$ & Uniform(0, 2$\pi$) \\
$\log_{10}(\mathcal{M}/\mathrm{M}_\odot)$ & Uniform(7, 10) \\
$\log_{10}(f_{\mathrm{GW}}$/Hz) & Uniform(-9, -7) \\
$\cos\iota$ & Uniform(-1, 1) \\
$\psi$ & Uniform(0, $\pi$) \\
$\Phi_0$ & Uniform(0, 2$\pi$) \\
$\log_{10}h$ & Uniform(-18, -11) \\
\cline{1-2}
$L_i$ & Normal($L_i$, $\sigma_{\mathrm{L_i}}$) \\
$\Phi_i$ & Uniform(0, 2$\pi$) \\
\enddata
\end{deluxetable}

\subsection{Galaxy Catalog} \label{sec:gxycat}

We use NANOGrav's catalog of massive galaxies in the local universe assembled in \cite{2021ApJ...914..121A}. Derived from the 2MASS Redshift Survey \citep[2MRS;][]{2012ApJS..199...26H}, it includes sky coordinates, distances, and SMBH masses for 43,532 galaxies out to redshift $z \sim 0.05$. The catalog is 97.6\% complete to $\sim 300$ Mpc and apparent $K$-band magnitudes $m_K \leq 11.75$, and it extends out to $\sim$ 500$-$700 Mpc for the most massive galaxies that will be prime targets for PTAs \citep[for a full scope of the SMBH masses and distances covered by the catalog, see Figure 8 in][]{2021ApJ...914..121A}. We note that while 2MRS is an all-sky survey, it avoids the galactic plane, i.e., galactic latitude $|b| < 5^{\circ}$, corresponding to about 9\% of the sky. See Figure 1 in \cite{2012ApJS..199...26H} for more details. Finally, the catalog includes important quantities for multi-messenger detections, like the distance of the galaxy and the mass of the central SMBH.

Throughout this work, we interpret each galaxy's SMBH mass to be the total mass of a putative binary rather than a single black hole mass. The SMBH masses are calculated using the most accurate method available for each galaxy, including dynamical measurements \citep[from observations of stellar, gaseous, or maser kinematics;][]{2013ApJ...764..184M,2013Natur.494..328D,2014Natur.513..398S,2014AAS...22312607W,2016ApJ...817....2W,2016Natur.532..340T}, reverberation mapping \citep{2015PASP..127...67B}, the $M_{\mathrm{BH}}-\sigma$ relation \citep{2013ApJ...764..184M}, and the $M_{\mathrm{BH}}-M_{\mathrm{bulge}}$ relation \citep{2013ApJ...764..184M}. The level of uncertainty in the SMBH mass varies across these methods; the most accurate masses come from dynamical measurements (0.01$-$0.33~dex uncertainty) and reverberation mapping (0.02$-$0.22~dex), followed by those estimated from the $M_{\mathrm{BH}}-\sigma$ relation (0.36$-$0.46~dex), and finally, those estimated from the $M_{\mathrm{BH}}-M_{\mathrm{bulge}}$ relation (0.4$-$0.48~dex).

A majority of the galaxies in the catalog have SMBH masses derived from the $M_{\mathrm{BH}}-M_{\mathrm{bulge}}$ relation. For these galaxies, the $K$-band luminosity is used to calculate the total stellar mass $M_*$, which is then used to estimate the bulge mass as $M_{\mathrm{bulge}} = f_{\mathrm{bulge}}M_*$, where $f_{\mathrm{bulge}}$ is the fraction of stellar mass residing in the bulge. However, about half of the galaxies in the catalog have unknown morphological types and, consequently, unknown $f_{\mathrm{bulge}}$ quantities, introducing additional uncertainty into the $M_{\mathrm{bulge}}$ calculation. Unknown-type galaxies therefore have two SMBH mass estimates in the catalog, which use two different values of bulge fraction; $f_{\mathrm{bulge}}=1.0$, corresponding to elliptical galaxies, and $f_{\mathrm{bulge}}=0.31$, corresponding to spiral Sa-type galaxies. For more information about the SMBH masses in the catalog, see \cite{2021ApJ...914..121A}. 
 
\subsection{Simulated PTA Datasets} \label{sec:sims}

\subsubsection{PTA Configuration}

We simulate realistic datasets similar to the in-preparation IPTA-DR3, including 116 pulsars timed across all of the constituent PTAs. The sky map in \autoref{fig:9gxy_snrhist} shows the distribution of pulsars in our simulated array, represented by white stars. Beginning with the 68 pulsars timed in the NANOGrav 15 year dataset \citep{2023ApJ...951L...9A}, we add on all non-NANOGrav pulsars in the following order, without repeating pulsars: 14 pulsars from PPTA DR3 \citep{2023arXiv230616230Z}, 3 pulsars from EPTA+InPTA DR2new+ \citep{2023arXiv230616214A}, and 31 pulsars from MPTA DR1 \citep{2023MNRAS.519.3976M}. This ordering ensures that any non-NANOGrav pulsar timed by multiple PTAs is added as part of the PTA in which it has the longest baseline\footnote{We note that there are some NANOGrav pulsars that have longer baselines in other PTAs. However, because we chose to build our simulated PTA as an extension of the NANOGrav PTA, we therefore use NANOGrav baselines for NANOGrav-timed pulsars.}. We use pulsar distance values listed in Table 2 of \cite{2023ApJ...951L..50A} for all NANOGrav pulsars, Table 3 of \cite{2023arXiv230616224A} for the 3 EPTA+InPTA DR2new+ pulsars, and the Australia Telescope National Facility Pulsar Catalogue\footnote{https://www.atnf.csiro.au/people/pulsar/psrcat} \citep{2005AJ....129.1993M} for all other pulsars.

To emulate the sensitivity of the NANOGrav 15-year dataset, we adopt the 12.5-year dataset \citep{2021ApJS..252....4A} as the foundation of our simulations. All 45 pulsars in this subset of the array have the same observational timestamps and TOA uncertainties as those in the real NANOGrav dataset. We then extend the timing baseline to 20 years, as this will be the approximate baseline of the IPTA-DR3 dataset. We employ the methods outlined in \cite{2021ApJ...911L..34P}, using the statistics of the last year of observations from the 12.5-year dataset to generate future observational data with a new cadence and new TOA uncertainties. To reasonably represent the sensitivity of these 45 pulsars following the collapse of the Arecibo Observatory, we assume double the observing time at the Green Bank Telescope, and for pulsars previously observed by Arecibo, the TOA uncertainties are inflated to reflect the change in the telescope's system equivalent flux density.

For the other 71 pulsars (those added on in the NANOGrav 15-year dataset, as well as the PPTA pulsars, EPTA+InPTA pulsars, and MPTA pulsars), we implement the following procedure. First, we assemble a group of reasonably unremarkable real pulsars to fabricate template timing models. More specifically, we choose pulsars that are not highly sensitive outliers, are not part of a binary system, and do not have any measured intrinsic red noise. We use the pulsars J0931-1902, J1453+1902, J1832-0836, and J1911+1347, which were added to the array between the NANOGrav 11-year dataset and 12.5-year dataset. For each of the 71 simulated pulsars, we randomly choose one of the ``average" template pulsars from which to adopt timing model parameters, replacing the template sky location with the true desired location.

These pulsars are timed over the baseline listed in their respective dataset papers, plus an additional $\sim4$~years to reach a 20-year baseline. Across all pulsars, the NANOGrav 15-year dataset spans roughly 16 years from the first observation to the last; therefore, to reach the 20-year mark, we only need an additional 4 years of observations for each pulsar. Note that this is a 20-year baseline for the NANOGrav array, whereas the overall baseline of the entire 116-pulsar array is $\sim$ 22 years. This is due to the fact that a handful of pulsars in other PTAs have observations preceding the first NANOGrav observation, making their individual baselines longer than 20 years. The observations for all 71 simulated pulsars are carried out every 2 weeks, and their TOA uncertainties are taken to be the whitened RMS values in the dataset papers.

We do not inject any white or red noise into these simulations but instead model these processes in the noise covariance matrix during the analysis stage. This approach allows us to effectively obtain the average posterior distributions over many noise realizations without actually generating entire suites of realizations \citep{2010ApJ...725..496N, 2010arXiv1007.4820C}. 

\subsubsection{Host Galaxies and Injected Signals}

\begin{figure*}[!ht]
    \centering \includegraphics[width=0.98\textwidth]{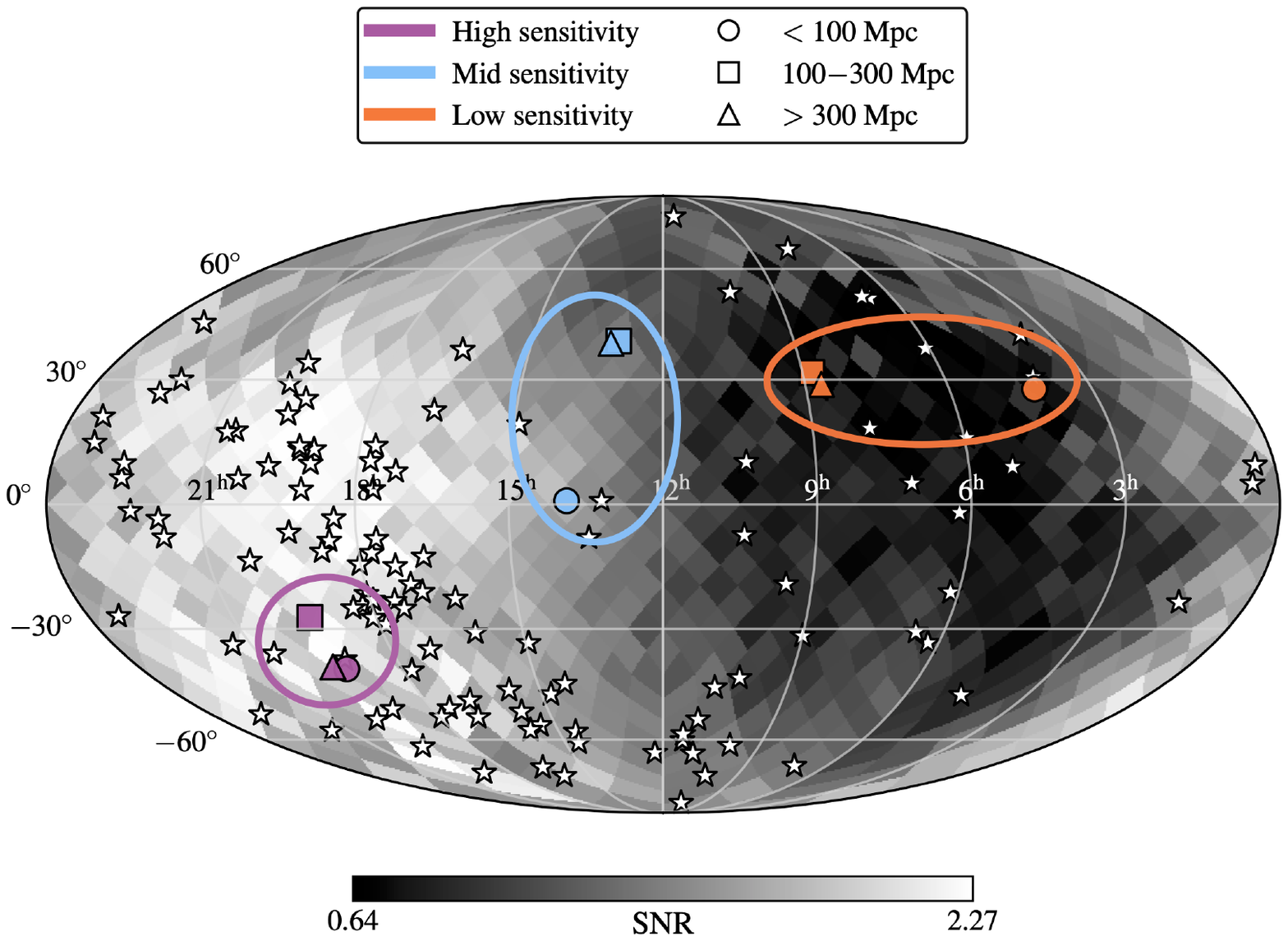}
    \caption{Skymap of the simulated PTA's sensitivity, where each pixel across the sky is colored by the expected SNR of a CW signal coming from that sky location, fixed at $d_L =$ 150 Mpc, and averaged across different binary parameter values (GW frequencies of 2 nHz, 6 nHz, and 20 nHz; chirp masses of $10^8$ $\mathrm{M}_\odot$ and $10^9$ $\mathrm{M}_\odot$; and 100 random draws of $\cos\iota$, $\psi$, and $\Phi_0$). Pulsar positions are represented by the white stars. We inject into a total of nine galaxies, grouped into high (purple markers), mid (light blue markers), and low (orange markers) sky sensitivity regions. These three groups are demarcated by ellipses of the same color. In each sensitivity region, we inject into one galaxy at each of the near, mid, and far distance regimes, corresponding to $<$ 100 Mpc (circles), 100$-$300 Mpc (squares), $>$ 300 Mpc (triangles), respectively.}
    \label{fig:9gxy_snrhist}
    \vspace{3 mm}
\end{figure*}

\autoref{fig:9gxy_snrhist} shows how our simulated PTA's sensitivity to single sources changes across the sky. The sky map is color-coded by the expected signal-to-noise ratio (SNR), which is calculated as the noise-weighted inner product:

\begin{equation}
\mathrm{SNR} = \langle \boldsymbol{s} | \boldsymbol{s} \rangle ^{1/2} = \sqrt{\boldsymbol{s}^T \cdot C^{-1} \cdot \boldsymbol{s}}
\end{equation}

\noindent where $\boldsymbol{s}$ is again the vector symbolizing the CW signal (now concatenated over all pulsars), and $C$ is the PTA noise covariance matrix containing the intrinsic white and red noise in each pulsar as well as the common red noise process used to model the GWB. Given that we model the GWB here as a common uncorrelated red noise process, $C$ is block diagonal in terms of pulsars. Our expected SNR skymap is computed for a CW signal injected into each of 768 equal-area pixels on the sky, at $d_L=150$ Mpc and averaged over GW frequencies of 2 nHz, 6 nHz, and 20 nHz, chirp masses of $10^8$~M$_{\odot}$ and $10^9$~M$_{\odot}$, and 100 random draws of the parameters $\cos\iota \in [-1,1]$, $\psi \in [0,\pi]$, and $\Phi_0 \in [0,2\pi]$. Our array's sensitivity closely follows the distribution of pulsars, with the highest SNR pixels being in the left half of the sky where a majority of the pulsars lie, and the lowest SNR pixels similarly appearing in the right half where there are fewer pulsars.

In \autoref{fig:9gxy_snrhist} we also mark the nine fiducial ``truth” galaxies used for this study. For each galaxy, we inject a CW signal, determine the localization area of the signal, and quantify the number of potential host galaxies within that localization area. When choosing our fiducial set of galaxies, we take a somewhat conservative approach and inject into those that have a SMBH mass determined only from the $M_{\mathrm{BH}}-M_{\mathrm{bulge}}$ relation (i.e., the galaxies that have the highest uncertainty in their mass estimates), as we expect the majority of galaxies to have this kind of estimate as opposed to one obtained using dynamical measurements.
 
We also choose only from galaxies that have known morphological types (and, therefore, known $f_{\mathrm{bulge}}$ values). Finally, because we want to focus on binaries that would be detectable by PTAs, we choose galaxies with a total SMBH mass $M_{\mathrm{BH}} \geq 10^9$ $\mathrm{M}_{\odot}$, which roughly translates to a stellar mass threshold of $M_* \geq 10^{11.5}$ $\mathrm{M}_{\odot}$.

From the 3325 galaxies in the catalog that fit these criteria, our nine fiducial galaxies are chosen somewhat arbitrarily, but on the condition that they are distributed across three sky regions of varied sensitivity as well as three distance shells. Three galaxies are located in the region of the sky that is most sensitive (purple markers, where there are many pulsars), three are in the least sensitive region (orange markers, where there are few pulsars), and three are in a region with intermediate sensitivity (light blue markers). For clarity, we include ellipses in \autoref{fig:9gxy_snrhist} to define these three groups. Each group of three contains one galaxy at a distance $<$ 100 Mpc, one between 100$-$300 Mpc, and one $>$ 300 Mpc away, represented by circle, square, and triangle markers, respectively.

The injected distances are taken from the galaxy catalog. The injected chirp mass is $\mathcal{M} = q^{3/5}/(1+q)^{6/5}M_{\mathrm{tot}}$, with the total binary mass $M_{\mathrm{tot}}$ taken as the SMBH mass from the galaxy catalog, while the mass ratio $q$ is adjusted in order to fix the desired SNR. We choose to tweak the mass ratio of the source rather than the distance, as changing the distance means that our hosts will no longer correspond to true galaxies from the catalog. We note that changing the mass ratio across the nine galaxies will also affect their chirp masses, and thus slightly influence the relative frequencies of the pulsar terms as well. In \autoref{tab:9table} we list the galaxy name (or sky coordinates), distance, and total SMBH mass for each galaxy, as well as the injected mass ratio for two sets of injections, one with fixed SNR=8 and the other with fixed SNR=15.

All 18 injections have a fixed GW frequency of $f_{\mathrm{GW}}=20$~nHz, face-on inclination angle of $\iota=0$, GW polarization angle $\Psi = \pi/4$, and initial binary phase $\Phi_0 = \pi/4$. The injected GW frequency and inclination angle are idealized choices. A frequency of 20 nHz lies in the ``bucket" of current PTA sensitivity curves; in this regime, the CW signal is not only easier to disentangle from the GWB that manifests at lower frequencies, but it is also not such a high frequency that it is rapidly evolving within the timespan of our dataset. A face-on inclination angle produces the strongest signal, as in \autoref{eq:s_+} and \autoref{eq:s_x}. Our GW polarization angle and initial binary phase choices are arbitrary, but we do not expect them to change the generality of our results.

\begin{deluxetable*}{ccccc}
\setlength{\tabcolsep}{2.25em}
\tablecaption{Injected CW parameters for the fiducial nine galaxies. From left to right, the columns are: 2MASS ID, luminosity distance, total SMBH mass, mass ratio for fixed SNR=8, and mass ratio for fixed SNR=15. All injections have a GW frequency $f_{\mathrm{GW}}$ = 20 nHz, face-on binary inclination angle $\iota=0$, GW polarization angle $\psi=\pi/4$, and initial binary phase $\Phi_0=\pi/4$.\label{tab:9table}}
\tablecolumns{5}
\tablehead{
\colhead{Galaxy ID} & \colhead{$d_L$ (Mpc)} & \colhead{$\log_{10}(M_{\mathrm{tot}}/\mathrm{M}_{\odot})$} & \colhead{$q_8$} & \colhead{$q_{15}$}
}
\startdata
J19163258$-$4012332&76.9&9.26&0.099&0.31 \\
J19231198$-$2709494&276.6&9.53&0.1103&0.2795 \\
J19332496$-$3940214&323.0&9.63&0.1082&0.305 \\
J13523589+0049058&73.7&9.26&0.1333&0.405 \\
J13010676+3950290&157.3&9.53&0.1228&0.32 \\
J13110866+3913365&314.3&9.70&0.1241&0.3452 \\
J04122834+2742065&55.9&9.31&0.104&0.25 \\
J08475906+3147083&289.6&9.75&0.15&0.388 \\
J08391582+2850389&339.3&9.91&0.0823&0.1927 \\
\enddata
\end{deluxetable*}

\subsection{Host Galaxy Cuts} \label{sec:cuts}

For each injection, we follow the procedure in Section~\ref{sec:model2} to obtain the posterior distributions of the CW parameters. To quantify the localization region, we take the following steps: First, we take the posterior samples on the sky location ($\cos\theta$ and $\phi$) and use the \texttt{healpy} package \citep[based on the HEALPix\footnote{\url{http://healpix.sf.net}} scheme;][]{Zonca2019, 2005ApJ...622..759G} to create a probability density map with equal-area sky pixels. Next, we sort the pixels in order of descending posterior probability density, and cumulatively sum the probability densities to assign a credible level to each pixel. Finally, we compute the area associated with each credible level by counting the number of pixels within the given level and multiplying that number by the area of a pixel. Because the resulting curve of localization areas is a function of discrete credible level values, we interpolate along the curve to estimate the size of the desired credible area. Going forward, we use the 90\% credible area.

The skymap resolution, or $N_{\mathrm{side}}$ value (dividing the sky into $12N_{\mathrm{side}}^2$ pixels), is not constant across all skymaps in this work. Rather, the resolution is chosen based on the source localization, with more tightly constrained sky posteriors necessitating higher resolution maps. For every skymap, we pick the resolution by varying the number of pixels from $N_{\mathrm{side}} = 8$ (768 pixels) to $N_{\mathrm{side}} = 64$ (49,152 pixels) and carrying out the procedure outlined above for each $N_{\mathrm{side}}$. We then compare the localization area curves obtained across $N_{\mathrm{side}}$ values and choose the $N_{\mathrm{side}}$ at which the curves begin to converge.

To estimate MCMC sampling uncertainties on the localization areas, we use the bootstrapping method of resampling with replacement. More specifically, we create 100 bootstrapped sets of posterior samples by random sampling with replacement and compute the localization area for each using the procedure described above. The confidence intervals are then assigned using the distribution of the computed areas. Throughout the paper, we quote the median localization areas, and all uncertainties correspond to the 68\% confidence interval.

The next step in the pipeline is to determine how many potential hosts from NANOGrav's galaxy catalog are enclosed within the 90\% credible area. This is done with \texttt{matplotlib} by counting the number of points lying within the boundary of the 90\% credible level contour. We note that, throughout this work, our localization areas are plotted with the default smoothing internal to \texttt{ligo.skymap}; therefore, when counting the number of hosts within a given area, we use the default-smoothed contours. The uncertainties on the localization areas inherently imply that the number of host galaxies enclosed within these areas, too, may be somewhat variable. However, for any given injection, we simply show the number of galaxies obtained for one representative bootstrapped sky location posterior. Further discussion of these choices and uncertainties on the number of galaxies can be found in Section \ref{sec:discuss_cuts}.

Finally, we implement cuts on the enclosed galaxies based on the chirp mass and luminosity distance posteriors recovered from the GW search. In order to directly compare against the total SMBH mass estimates in the galaxy catalog, we decompose our chirp mass posterior into distributions of the total binary mass and mass ratio using rejection sampling. First, our chirp mass posterior serves as a target distribution from which we resample in total mass and mass ratio. We propose pairs of ($\log_{10}M_{\mathrm{tot}}$, $\log_{10}q$) from uniform priors $\log_{10}(M_{\mathrm{tot}}/\mathrm{M}_{\odot}) \in [8,11]$ and $\log_{10}q \in [-1,0]$ and calculate the corresponding chirp mass $\log_{10}\mathcal{M}$. We then sample a uniform random number between zero and the maximum of the target $\log_{10}\mathcal{M}$ distribution. If this uniform random draw is smaller than the target distribution evaluated at the proposed chirp mass, then the proposed ($\log_{10}M_{\mathrm{tot}}$, $\log_{10}q$) values are accepted. This procedure is repeated to generate draws from our new joint posterior on $\log_{10}M_{\mathrm{tot}}$ and $\log_{10}q$.

We test the level of agreement between the total binary mass posterior and the total SMBH mass for any given galaxy with the non-Gaussian tension estimator \texttt{tensiometer} \citep{2021PhRvD.104d3504R}. The package first draws one sample from each distribution and computes the difference, repeating this process until the difference distribution is thoroughly sampled. The probability that the difference is zero is determined by integrating this distribution, and the Gaussian-equivalent standard deviation corresponding to this probability is the final tension estimate.

\begin{figure}[!ht]
    \centering
    \includegraphics[width=0.45\textwidth]{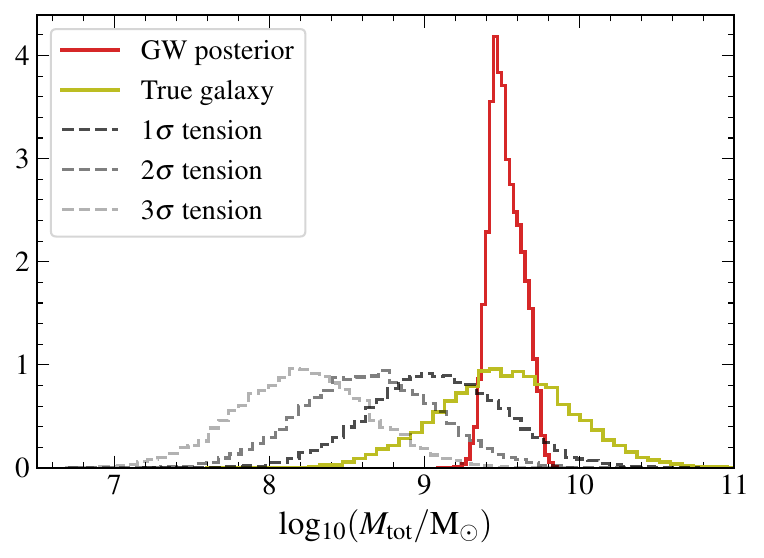}
    \caption{Tension metric examples as computed with \texttt{tensiometer}. The total binary mass posterior for the SNR=15 injection into galaxy J19231198$-$2709494 is shown in red. The true galaxy's total SMBH mass estimate (derived from the $M_{\mathrm{BH}}-M_{\mathrm{bulge}}$ relation), with a tension of $0.04\sigma$, is shown in yellow. Example distributions with $1\sigma$, $2\sigma$, and $3\sigma$ tension with the posterior are indicated by successively lighter-colored dashed lines.}
    \label{fig:tens_example}
\end{figure}

In \autoref{fig:tens_example} we show an example posterior distribution of the total binary mass for an SNR=15 injection, as well as the SMBH mass estimate of the true galaxy. The tension between these two distributions is $0.04\sigma$. For comparison, we include three distributions in $1\sigma$, $2\sigma$, and $3\sigma$ tension with the GW posterior. The tension values will have some dependence on the SMBH mass uncertainties from the galaxy catalog; for example, SMBH masses determined with dynamical measurements may be in higher tension with the GW posterior as compared to SMBH masses calculated from the $M_{\mathrm{BH}}-M_{\mathrm{bulge}}$ relation. Ideally, though, the true galaxy's SMBH mass would have low tension with the GW posterior.

The SMBH mass tension is calculated for each galaxy enclosed in a given 90\% credible localization area. From the distribution of tension values, we choose a reasonable threshold applicable to all injections and discard any galaxies above this tension threshold. We use example threshold values of $1\sigma$ and $2\sigma$, which we discuss in more detail in Section \ref{sec:results_cuts}. In the final step, we use the posterior on the luminosity distance to implement an additional cut on the remaining galaxies. We find the $95^\mathrm{th}$ percentile of the posterior and discard any potential hosts with distances beyond this value. Other distance cut options are briefly discussed in Section~\ref{sec:discuss_cuts}. Distance uncertainties are not considered as they are insignificant in comparison to SMBH mass uncertainties.

\section{Results} \label{sec:results}

\subsection{GW Source Localization} \label{sec:results_localization}

\begin{figure*}[!ht]
    \includegraphics[width=\textwidth]{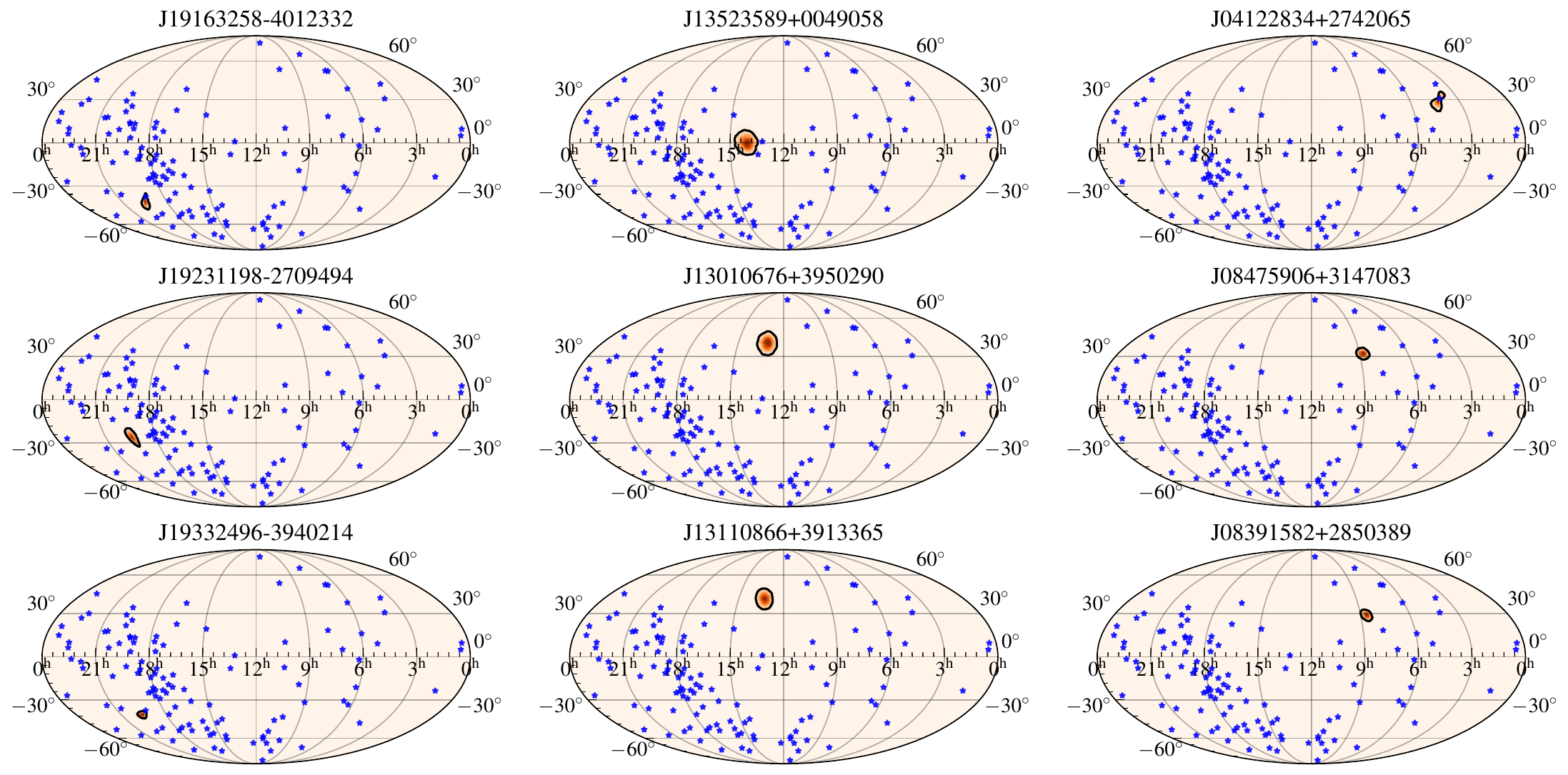}
    \caption{Skymaps of the 90\% credible level localization areas recovered for the SNR=15 injections, shown by the orange two-dimensional histograms outlined by black contours. The host galaxy IDs for each injection are listed at the top of each map. From left to right, the columns correspond to signals localized in regions of the sky with high, mid, and low sensitivity. From top to bottom, the rows correspond to injections with increasing distance ranges of $d_L <$ 100 Mpc, 100 Mpc $< d_L <$ 300 Mpc, and $d_L >$ 300 Mpc. The rows also roughly correspond to different chirp mass ranges, with the lowest chirp masses in the top row and increasing with each successive row. The blue stars represent the pulsar positions.}
    \label{fig:9gxy_snr15}
    \vspace{3 mm}
\end{figure*}

In the first set of simulations, we injected each of our nine fiducial galaxies with a CW signal of SNR=15. The recovered 90\% credible regions for these injections are shown in \autoref{fig:9gxy_snr15}. Each column corresponds to each of the three sensitivity regions into which the signals were injected: injections into the high sensitivity region are in the left column, followed by the mid sensitivity region in the middle column, and the least sensitive region in the right column. The ``sensitivity regions" here refer to the sensitivity of our PTA as established in \autoref{fig:9gxy_snrhist}, where higher sensitivity corresponds to higher SNR for a fixed set of binary parameters. The rows in \autoref{fig:9gxy_snr15} correspond to the different distance ranges: sources with $ d_L <$ 100 Mpc are in the top row, 100 Mpc $\leq d_L < 300 $ Mpc in the middle row, and $d_L \geq 300$ Mpc in the bottom row. However, as all nine sources are fixed at the same SNR, the rows also roughly correspond to different chirp mass ranges -- increasing from the top down -- to account for the change in distance. We see that for all nine cases, the signal is well-localized, with the 90\% credible region ranging from $\sim$ 29 deg$^2$ (bottom left) to $\sim$ 241 deg$^2$ (top center). Four sources, all three in the high sensitivity region (left column) and one in the low sensitivity region (top right), lie close to a pulsar on the sky, giving rise to the characteristic lobe-shaped areas that reflect the pulsars' antenna response pattern.

\begin{figure*}[!ht]
    \includegraphics[width=\textwidth]{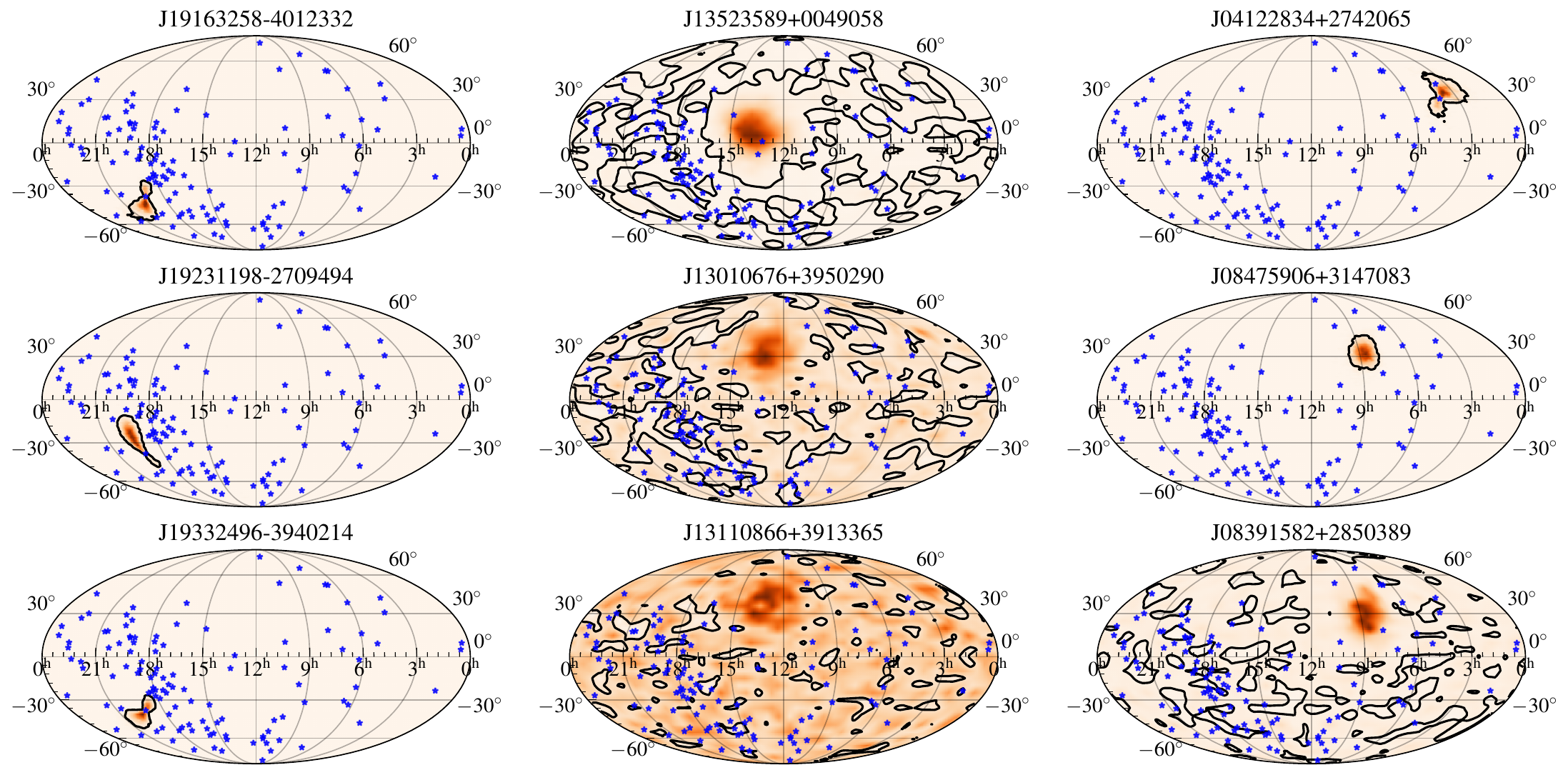}
    \caption{Skymaps of the 90\% credible level localization areas recovered for the SNR=8 injections, organized in the same manner as \autoref{fig:9gxy_snr15}. Note that the black contours enclose all regions included in the localization area, except for the middle center, bottom center, and bottom right skymaps, where the contours instead indicate excluded regions.}
    \label{fig:9gxy_snr8}
    \vspace{3 mm}
\end{figure*}

To get a more realistic idea of the scale of localization areas that PTAs may face when a single source is first detected, we also injected our nine galaxies with signals of SNR=8. We note that while an individually resolvable binary may be detected at lower SNRs, studying parameter estimation (particularly the localization) requires a higher SNR than that needed for a detection. For this reason, we choose an SNR=8 as our weak signal scenario. All CW parameters in this set are the same as in the SNR=15 injections apart from the mass ratio, which is adjusted to keep the SNR constant. By changing the mass ratio, the chirp masses and relative pulsar term frequencies in the SNR=8 set will also necessarily be different from those in the SNR=15 set. The 90\% credible areas are shown in \autoref{fig:9gxy_snr8}. 

The posteriors of the SNR=8 sources span much wider areas, as is expected for weaker signals. Within this set, five sources are well-localized, with areas ranging from $\sim$ 287 deg$^2$ (bottom left) to $\sim$ 530 deg$^2$ (top right). Compared to the areas obtained for the same sources in the SNR=15 set, these five areas are approximately 5 times larger at best and 10 times larger at worst. The sky locations of the other four SNR=8 sources are not recovered well enough to be informative, in that the posterior distributions on $\cos\theta$ and $\phi$ fill a large fraction of the prior. The localization areas for these sources range from $\sim$ 19,400 deg$^2$ (top center) to $\sim$ 32,400 deg$^2$ (bottom center), covering about 47\% to 79\% of the sky. Until the GW signal can be better localized, conducting any sort of host galaxy searches in such cases would be impractical.

We performed two additional sets of simulations for the fiducial nine galaxies, injecting CW signals with more conservative choices of the binary inclination angle and GW frequency. In one set, we kept the GW frequency at $f_{\mathrm{GW}} = 20 $ nHz and SNR=8 but instead chose an intermediate inclination angle of $\iota=\pi/3$. Since this inclination angle produces a weaker signal, keeping the SNR fixed means that the chirp masses are slightly larger in this set compared to those in the face-on SNR=8 set. However, the change in inclination angle did not significantly affect the recovered localization areas. Similar to those in \autoref{fig:9gxy_snr8}, some sources were well-localized while others were completely unconstrained. In the other set, we kept a face-on inclination angle of $\iota=0$ while modifying the GW frequency to be $f_{\mathrm{GW}}=6$~nHz, where the GWB is more prominent. When adjusting the mass ratio to obtain the desired SNR, the maximum value $q = 1$ did not yield an SNR $>$ 7 for all injections in this set; therefore, our $f_{\mathrm{GW}} = 6 $ nHz injections were fixed at SNR=7 rather than SNR=8. At such a low SNR, most of the sources in this group were unconstrained, indicating that an SNR $\sim$ 8 is needed for sufficient localization. We briefly discuss this finding in the next section, and further investigation of less-than-ideal parameter choices is left for another work.

\subsection{Factors Contributing to Localization}\label{sec:results_areafac}

Despite the SNR being fixed across all nine sources, the size of the localization area varies considerably in both sets of injections. We find that the source localization may depend on several factors, with the SNR being of initial consideration, followed by the source's proximity to pulsars on the sky, and lastly its chirp mass being of ``higher-order" importance.

\subsubsection{Signal-to-Noise Ratio}

As the area is known to scale as $1/\mathrm{SNR}^2$ \citep{2010PhRvD..81j4008S}, we begin by taking the three galaxies in the 100$-$300~Mpc distance range, each lying in a different region of the sky, and inject into them signals of varying SNR. Again, to adjust the SNR, we change the mass ratio (and therefore the chirp mass) of the binary. All of the injected mass ratios are listed in \autoref{tab:SNRtable}, with the first row corresponding to the lowest SNRs.

\begin{deluxetable}{ccc}
\setlength{\tabcolsep}{0.75em}
\tablecaption{Injected mass ratios for the SNR case study. From the top to bottom row, the mass ratios correspond to SNRs ranging from about 8 to 20. The injections are done for the three galaxies in the 100$-$300 Mpc distance group, with the high, mid, and low sensitivity hosts being J19231198$-$2709494, J13010676+3950290, and J08475906+3147083, respectively. All have fixed parameters $f_{\mathrm{GW}}=20$ nHz, $\iota=0$, $\psi=\pi/4$, and $\Phi_0=\pi/4$.\label{tab:SNRtable}}
\tablehead{
\colhead{High sensitivity} & \colhead{Mid sensitivity} & \colhead{Low sensitivity}
}
\startdata
0.1133 & 0.1442 & 0.1473 \\
0.1666 & 0.1740 & 0.1758 \\
0.2200 & 0.2112 & 0.2218 \\
0.2733 & 0.2781 & 0.2796 \\
0.3266 & 0.3451 & 0.2962 \\
0.3800 & 0.4120 & 0.3816 \\
0.4333 & 0.4790 & 0.5939 \\
0.4866 & 0.5587 & 0.6684 \\
0.5400 & 0.6798 & 0.6971 \\
\enddata
\end{deluxetable}

\begin{figure}[!ht]
    \centering
    \includegraphics[width=0.45\textwidth]{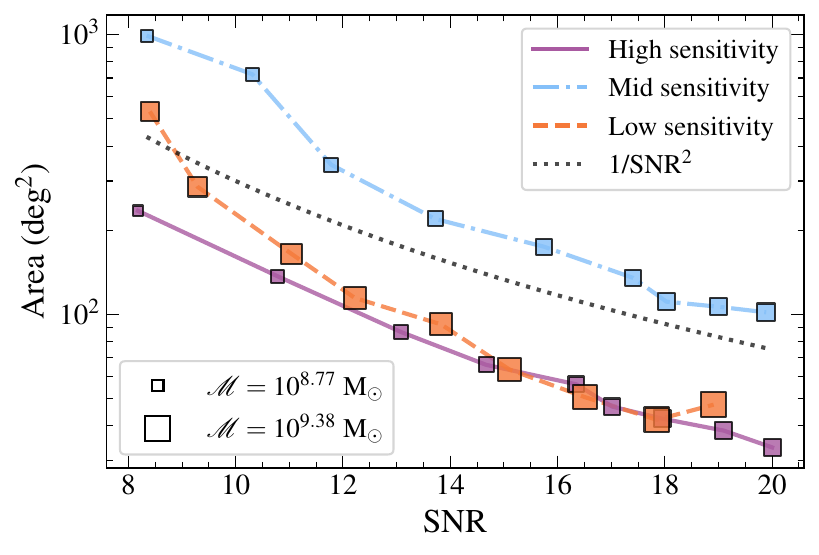}
    \caption{90\% credible area as a function of the injected SNR. The solid purple, dash-dotted light blue, and dashed orange curves correspond to each of the three galaxy hosts in the 100$-$300 Mpc distance range, located in the high (J19231198$-$2709494), mid (J13010676+3950290), and low (J08475906+3147083) sensitivity regions of the sky, respectively. Uncertainties on the localization areas are $\sim$ 13 deg$^2$ at most (at the lowest SNR) and therefore do not appear in the figure. The marker sizes reflect the injected chirp masses, with larger markers indicating higher chirp masses. The dotted gray line shows the expected 1/SNR$^2$ scaling relation.}
    \label{fig:area_vs_snr}
\end{figure}

Although we injected SNRs roughly ranging from 5 to 20, we do not include localization areas for injections with SNR $\lesssim$ 8, as these areas were poorly constrained. We note that it is roughly around this SNR that the localization area rapidly declines, separating from the full range of the prior. This behavior has been seen for similar SNRs in other studies \citep{2010PhRvD..81j4008S,2016ApJ...817...70T,2018MNRAS.477.5447G,2019MNRAS.485..248G}, reiterating that the data generally become informative and reasonable localization can be achieved around a threshold SNR $\gtrsim$ 8.

The resulting 90\% credible areas are shown as a function of SNR in \autoref{fig:area_vs_snr}. The sources are colored according to the region of the sky in which they lie, and the injected chirp masses are represented by the different marker sizes, increasing from left to right. Some slight deviations aside, all three sources generally follow the 1/SNR$^2$ scaling, shown by the dotted gray line for reference. Between the three sources, however, there are two notable features: $(i)$ The source lying in the mid sensitivity region has significantly larger localization areas as compared to the high and low sensitivity regions; $(ii)$ The localization areas of the high and low sensitivity sources seem to converge for half of the injected SNRs. We explore these features in more detail throughout the remainder of this section.

\subsubsection{Proximity of Nearby Pulsars}

\begin{figure}[!ht]
    \centering
    \includegraphics[width=0.45\textwidth]{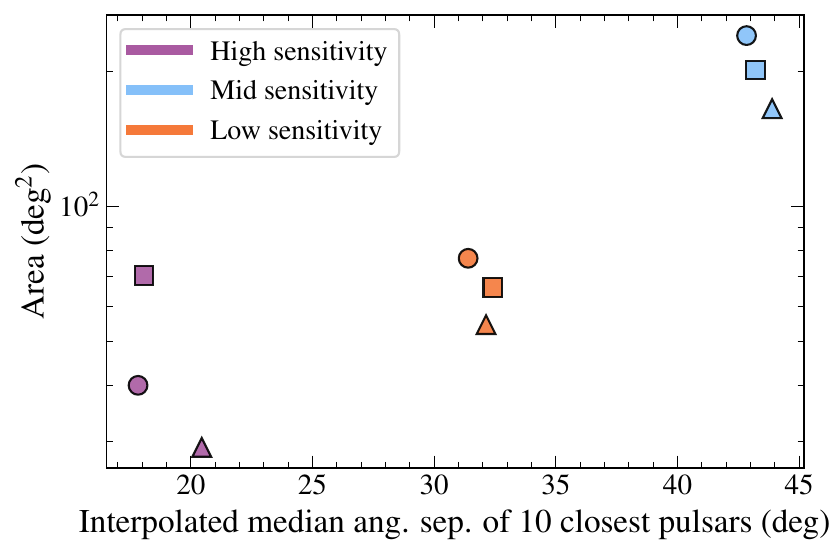}
    \caption{90\% credible area for each of the SNR=15 injections, shown as a function of the interpolated median angular separation between the source and the 10 nearest pulsars on the sky. Uncertainties on the localization areas are $\sim$ 2 deg$^2$ at most. As in \autoref{fig:9gxy_snrhist}, the sources' colors correspond to the different sensitivity regions, and their marker shapes correspond to the different distance ranges.}
    \label{fig:area_vs_psrangsep}
\end{figure}

First, the region with intermediate sensitivity has unexpectedly larger areas than those in the least sensitive region of the sky. While this feature is generally seen in \autoref{fig:9gxy_snr15} and \autoref{fig:9gxy_snr8} for fixed SNR, \autoref{fig:area_vs_snr} shows that the mid sensitivity region has the poorest localization regardless of the injected SNR. We see that this portion of the sky has a marked deficiency of pulsars, indicating that the proximity of pulsars' sky locations to that of the source may play an important role in its localization. To investigate this behavior further, we calculate the interpolated median angular separation of the ten closest pulsars to each of the nine fiducial galaxies and plot these separations against the SNR=15 localization areas in \autoref{fig:area_vs_psrangsep}.

Colored according to their sky locations, the sources fall along a positive relationship between the angular separation of nearby pulsars and the recovered localization area. Sources injected into the high sensitivity region naturally have the smallest areas. The high sensitivity of the PTA in this region is again due to the fact that there is an abundance of pulsars in this part of the sky, which looks towards the galactic center. All three sources injected here are consequently close to many pulsars, making the median separation to the ten nearest pulsars lower than that of the other sources.

While the sources in the low sensitivity region of the sky are generally expected to have the worst, or largest, localization areas, they substantially outperform the areas seen in the middle portion of the sky. We see that all three sources in the low sensitivity region have median separations from the nearest pulsars that are lower than those in the middle region, which has fewer nearby pulsars. These results are similar to the behavior seen in \cite{2019MNRAS.485..248G}, which we discuss more thoroughly in Section \ref{sec:discuss_comp}. We also note that this trend generally persists for different choices of the number of pulsars, as well as for other kinds of statistics, such as the mean of the nearest pulsars.

However, if we take the angular separations between each GW source in \autoref{fig:area_vs_psrangsep} and all 116 pulsars in the array, we instead see the expected trend: the median separation is lowest for the high sensitivity sources, followed by the mid sensitivity sources, and lastly the low sensitivity sources. Although the middle portion of the sky is less dense in pulsars, it is closer to all pulsars in the array on average, making it more sensitive to single sources. By contrast, the low sensitivity region has a higher density of pulsars but is generally farther away from all pulsars and consequently less sensitive to single sources. In other words, the density of pulsars on the sky is important for localization precision, but the sensitivity of a given sky region is determined by the entire PTA.

We choose not to include the SNR=8 localization areas in \autoref{fig:area_vs_psrangsep} as not all the sources from that set were well-constrained, particularly those in the mid sensitivity region. However, because the areas recovered for the mid sensitivity sources spanned large fractions of the sky, they provide further evidence of the importance of nearby pulsars. If additional pulsars in this region are identified and regularly timed, perhaps these areas could be better localized for weaker signals.

The closeness of nearby pulsars to a given GW source therefore has significant impact on a PTA's ability to localize the signal, regardless of the ``sensitivity region" in which the source is located. These results reflect the findings presented in \cite{2012PhRvD..86l4028B}, which put forth that the localization of the GW source is primarily governed by those pulsars which are closest to it on the sky. More specifically, when pulsar distances are not accurately known, the localization is determined by the smallest quadrilateral of pulsars around the source, i.e., the four pulsars whose positions make up the smallest sky area while still enclosing the source. Our simulated PTA roughly reflects this case, as the pulsar distance uncertainties in our array range from 1 pc (0.6\%) at best to 5 kpc (20\%) at worst. In order to be categorized as pulsars with accurately known distances, these uncertainties must be smaller than the gravitational wavelength, or on the order of $\lesssim$ 0.5 pc for our injected 20 nHz signals. While pulsar distances are difficult to measure, more precise measurements with uncertainties on this scale would be significant steps forward in improving the localization of CW signals in the future \citep{2012PhRvD..86l4028B, 2023PhRvD.108l3535K}. However, the contribution of nearby pulsars does not simply hinge on their proximity to the source and their distance uncertainties, but will also depend on other pulsar characteristics, such as their individual timing baselines and noise properties.

\subsubsection{Chirp Mass}

We now turn to the second feature seen in \autoref{fig:area_vs_snr} and compare the localization areas of the high and low sensitivity sources. The low sensitivity source generally does worse than the high sensitivity source, but the performance changes at SNR $\sim$ 15, after which the two have similarly-sized areas. This behavior may be due to the fact that the low sensitivity source has a total SMBH mass of $\log_{10}(\mathrm{M_{tot}}/\mathrm{M}_\odot)$ = 9.75, whereas the high sensitivity source has a total SMBH mass of $\log_{10}(\mathrm{M_{tot}}/\mathrm{M}_\odot)$ = 9.53. Therefore, when adjusting the mass ratio to inject the desired SNR, the injected chirp mass in the low sensitivity region is always higher than that in the high sensitivity region. We depict this variation in chirp mass through the marker sizes in \autoref{fig:area_vs_snr}, with the largest markers corresponding to the highest chirp masses. The source in the low sensitivity region clearly boasts higher chirp masses across all SNRs, yet the similarity in localization areas appears only for SNR $\gtrsim$ 15.

As shown in \autoref{eq:domega_dt}, the chirp mass is an important factor in the frequency evolution of the binary. Binaries with higher chirp masses will evolve more quickly, resulting in a greater difference between the earth term and pulsar term frequencies. While including the pulsar term in the signal model has been shown to improve constraints on the distance and chirp mass of the source in comparison to the earth term signal alone (in which case these parameters are highly degenerate within the overall signal amplitude), the pulsar term additionally allows for a more precise measurement of the binary position \citep{2010arXiv1008.1782C,2011MNRAS.414.3251L}. Indeed, from \autoref{eq:t_p} we see that the timestamp of the pulsar term is related to the angle between the pulsar’s position and the source’s position on the sky. Thus, as long as the Earth term and pulsar term frequencies can be disentangled from one another, we can expect the chirp mass to provide some additional constraint on the binary's sky location.

In order to isolate this effect on the localization area, we inject signals of varying chirp mass into the galaxy J08475906+3147083 (see the orange square in \autoref{fig:9gxy_snrhist}). To test the assumption that the pulsar term provides valuable information in localizing the source, we generate a second set of injections containing only the Earth term component of the signal; the analyses performed on these Earth-term-only injections are then done with Earth-term-only models comprised of just eight CW parameters. For both groups, we adjust the distance of the binary to keep the signal strength fixed, which we set to SNR=20 to ensure that all injections will be well-localized. \autoref{tab:MAtable} presents a complete list of chirp masses and distances for this case study's injections, all of which have parameters $f_{\mathrm{GW}}=20$ nHz, $\iota=0$, $\psi=\pi/4$, and $\Phi_0=\pi/4$.

\begin{deluxetable}{ccc}
\setlength{\tabcolsep}{0.75em}
\tablecaption{Injected parameters for the chirp mass case study, including the chirp masses and the distances for both the full signal ($d_{L\mathrm{,full}}$) and Earth-term-only ($d_{L\mathrm{,ETO}}$) injections. All injections are done for galaxy J08475906+3147083 and have parameters SNR=20, $f_\mathrm{GW}=20$ nHz, $\iota=0$, $\psi=\pi/4$, and $\Phi_0=\pi/4$.\label{tab:MAtable}}
\tablecolumns{2}
\tablehead{
\colhead{$\log_{10}(\mathcal{M}/M_{\odot})$} & \colhead{$d_{L\mathrm{,full}}$ (Mpc)} & \colhead{$d_{L\mathrm{,ETO}}$ (Mpc)}
}
\startdata
9.00 & 60 & 44 \\
9.05 & 64 & 53 \\
9.10 & 83 & 64 \\
9.15 & 111 & 78 \\
9.20 & 125 & 94 \\
9.25 & 153 & 114 \\
9.30 & 194 & 138 \\
9.35 & 236 & 168 \\
9.40 & 268 & 203 \\
9.45 & 341 & 246 \\
9.50 & 379 & 298 \\
\enddata
\end{deluxetable}

\begin{figure}[!ht]
    \centering
    \includegraphics[width=0.45\textwidth]{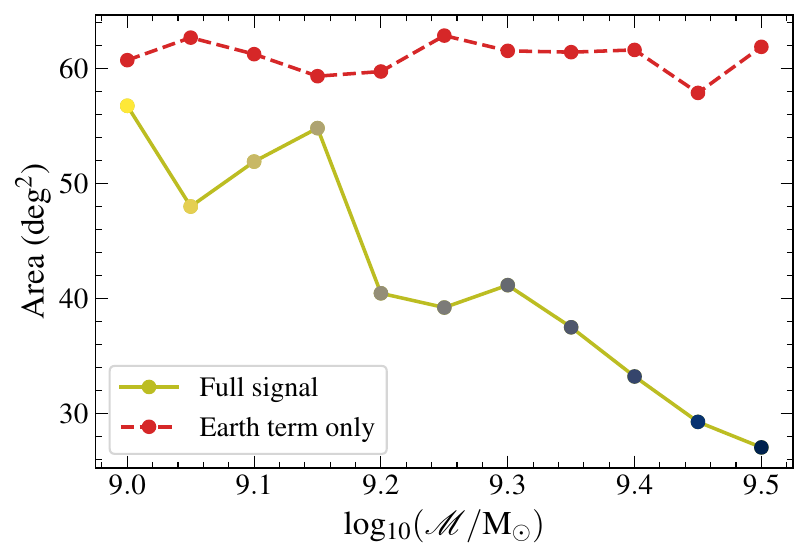}
    \caption{90\% credible area as a function of the injected chirp mass of the binary, fixed at SNR=20 for the galaxy J08475906+3147083. The full signal injections are shown by the solid yellow line, with lighter (or more yellow) markers representing lower chirp masses and darker (or more blue) markers representing higher chirp masses. The Earth-term-only injections are shown by the dashed red line. Uncertainties on the localization areas are all $<$ 1 deg$^2$.}
    \label{fig:area_vs_mc}
\end{figure}

The recovered 90\% credible areas for both the full and Earth-term-only signals are plotted as a function of the chirp mass in \autoref{fig:area_vs_mc}. Immediately, we see a stark difference between the two curves. The Earth-term-only signals, though not identical across all chirp masses, tend to hover around an average localization area of $\sim$ 61 deg$^2$. On the contrary, the full signals have consistently better-constrained sky areas than the Earth-term-only signals, and they reveal an entirely different pattern across the injected chirp masses. These areas initially exhibit some oscillatory behavior up to a chirp mass of about $\log_{10}(\mathcal{M}/\mathrm{M}_\odot) = 9.3 - 9.35$, at which point the area then decreases monotonically with increasing chirp mass. Both features can be attributed to the influence of the pulsar term.

At lower chirp masses, the fluctuation in localization area is likely due to the minimal frequency evolution of the source. In this scenario, the Earth and pulsar term frequencies are not sufficiently separate from one another in that they lie in the same frequency bin. Although CW searches are not bound to Fourier-bin resolution like GWB searches are, we use the frequency bins as reasonable metrics to assess the evolution of the binary. The frequency bins in our array are defined by $f_i = i/T$, where $T$ is the $\sim$ 22 year timespan of our dataset, which translates to bin widths of about $1.43$~nHz. For any pulsar-term frequency falling in the 18.59$-$20.02~nHz bin along with the injected Earth term frequency of $20$~nHz, the full signal will essentially become a sum of two sinusoids with different phases, which may constructively or destructively interfere \citep{2011MNRAS.414.3251L}. Of course, not every pulsar term will lie in the same frequency bin as the Earth term, since the difference between the two frequencies will depend on the pulsar's distance and projected angular separation from the source.

\begin{figure}[!ht]
    \centering
    \includegraphics[width=0.45\textwidth]{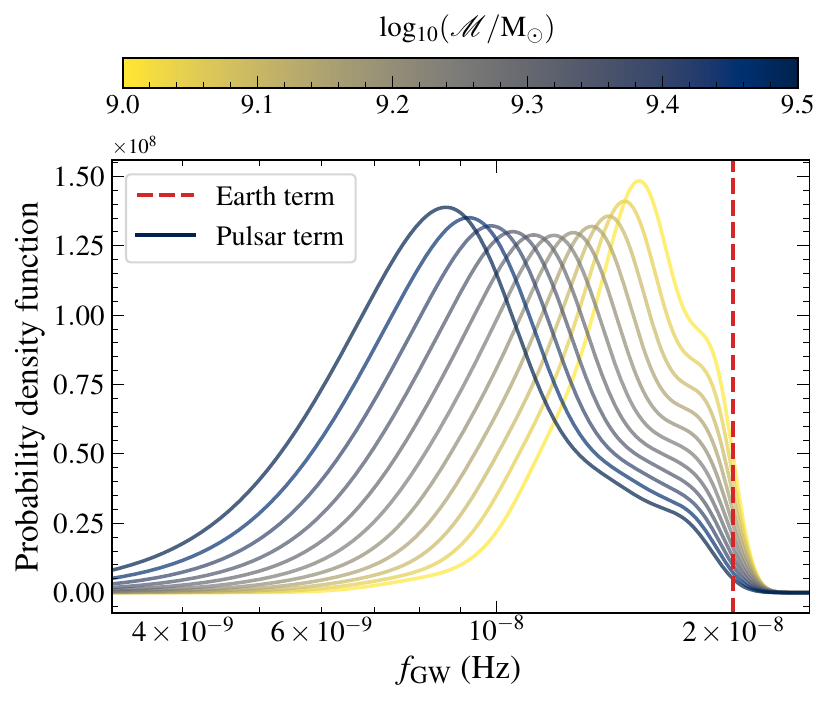}
    \caption{Distribution of pulsar term frequencies for the injected binary chirp masses in \autoref{fig:area_vs_mc}. The color scale corresponding to the different chirp masses is the same as in \autoref{fig:area_vs_mc}. The injected Earth term frequency of 20 nHz is indicated by the dashed red line.}
    \label{fig:pterm_freqs}
\end{figure}

In \autoref{fig:pterm_freqs}, we show more clearly the distribution of pulsar term frequencies for each full signal injection from \autoref{fig:area_vs_mc}. The distributions are colored by the injected chirp mass, with yellow corresponding to the lowest chirp mass of $\log_{10}(\mathcal{M}/\mathrm{M}_\odot)$ = 9.0 and progressively darker (or more blue) colors corresponding to higher chirp masses. The dashed red line marks the injected Earth term frequency of 20 nHz. Here we see that, for higher chirp masses, the distribution of pulsar term frequencies is pushed farther away from the Earth term frequency, as anticipated. We note that for the transition point in \autoref{fig:area_vs_mc}---chirp masses around $\log_{10}(\mathcal{M}/\mathrm{M}_\odot) = 9.3 - 9.35$ where the oscillating behavior ends and the monotonic decrease begins---the median pulsar term frequency is around 10.97$-$11.61~nHz.

Returning to \autoref{fig:area_vs_snr}, recall that the convergence of localization areas between the high and low sensitivity regions occurs at SNR $\sim$ 15. At this SNR, the injection for the low sensitivity source has a chirp mass of $\log_{10}(\mathcal{M}/\mathrm{M}_\odot) \sim 9.33$ and median pulsar term frequency of 11.22 nHz, similar to the thresholds found in \autoref{fig:area_vs_mc} and \autoref{fig:pterm_freqs}. By contrast, the chirp masses of the high sensitivity region injections do not exceed $\log_{10}(\mathcal{M}/\mathrm{M}_\odot) \sim 9.14$. Therefore, the similarity in localization areas around SNR $\sim$ 15 occurs because this is where the low sensitivity source's chirp mass is high enough to contribute to its localization, pulling the areas down to the same level as those of the high sensitivity source.

\subsubsection{Summary}

\autoref{fig:area_vs_snr} shows us that the SNR certainly provides the backbone scaling relation for the localization area of a given CW source, but the localization capability of a PTA can be further enhanced with the help of nearby pulsars and a high enough binary chirp mass. In \autoref{fig:area_vs_psrangsep}, the organization of the three sensitivity regions suggests that the recovered localization area for a source is largely dependent on the angular distance to the closest pulsars, due to the array’s anisotropic distribution on the sky. In \autoref{fig:area_vs_mc} and \autoref{fig:pterm_freqs}, at high enough chirp masses where there is substantial binary evolution, the difference between the earth term and pulsar term frequencies provides further information in constraining the source’s sky location.

All of these effects come together in \autoref{fig:9gxy_snr15} and \autoref{fig:9gxy_snr8}. The scale of the localization area is clearly different when comparing injections of SNR=15 in \autoref{fig:9gxy_snr15} to SNR=8 in \autoref{fig:9gxy_snr8}. However, the variety of areas within both sets of injections shows that the SNR is not the sole determinant, but is rather accompanied by contributions from nearby pulsars and high binary chirp masses. In \autoref{fig:9gxy_snr15}, the region with intermediate sensitivity (middle column) has unexpectedly larger areas than those in the least sensitive region of the sky (right column), highlighting the importance of the GW source’s proximity to pulsars on the sky. This is similarly seen in \autoref{fig:9gxy_snr8}, with the exception of one unconstrained skymap in the low sensitivity region (bottom right). The localization area in this skymap and that of the skymap directly above it (middle right) both lie in a position bereft of nearby pulsars and therefore lack support in constraining the localization area. However, we see that one source is well-localized while the other is not; since both sources have a chirp mass $\log_{10}(\mathcal{M}/\mathrm{M}_\odot) \lesssim$~9.3, it seems that their localization may be in the oscillatory phase seen in \autoref{fig:area_vs_mc}. With such low chirp masses, the evolution between the earth and pulsar terms is not significant, and the PTA's localization capability fluctuates. \autoref{fig:9gxy_snr15}, though, demonstrates the case in which the chirp mass does contribute to the localization. The middle right source has a chirp mass of $\log_{10}(\mathcal{M}/\mathrm{M}_\odot)$~=~9.33 and the bottom right source has a chirp mass of $\log_{10}(\mathcal{M}/\mathrm{M}_\odot)$ = 9.39, and their recovered localization areas are $\sim$ 66 deg$^2$ and $\sim$ 55 deg$^2$, respectively.

\subsection{Potential Hosts Remaining After Cuts} \label{sec:results_cuts} 

\begin{figure*}[!h]
    \includegraphics[width=\textwidth]{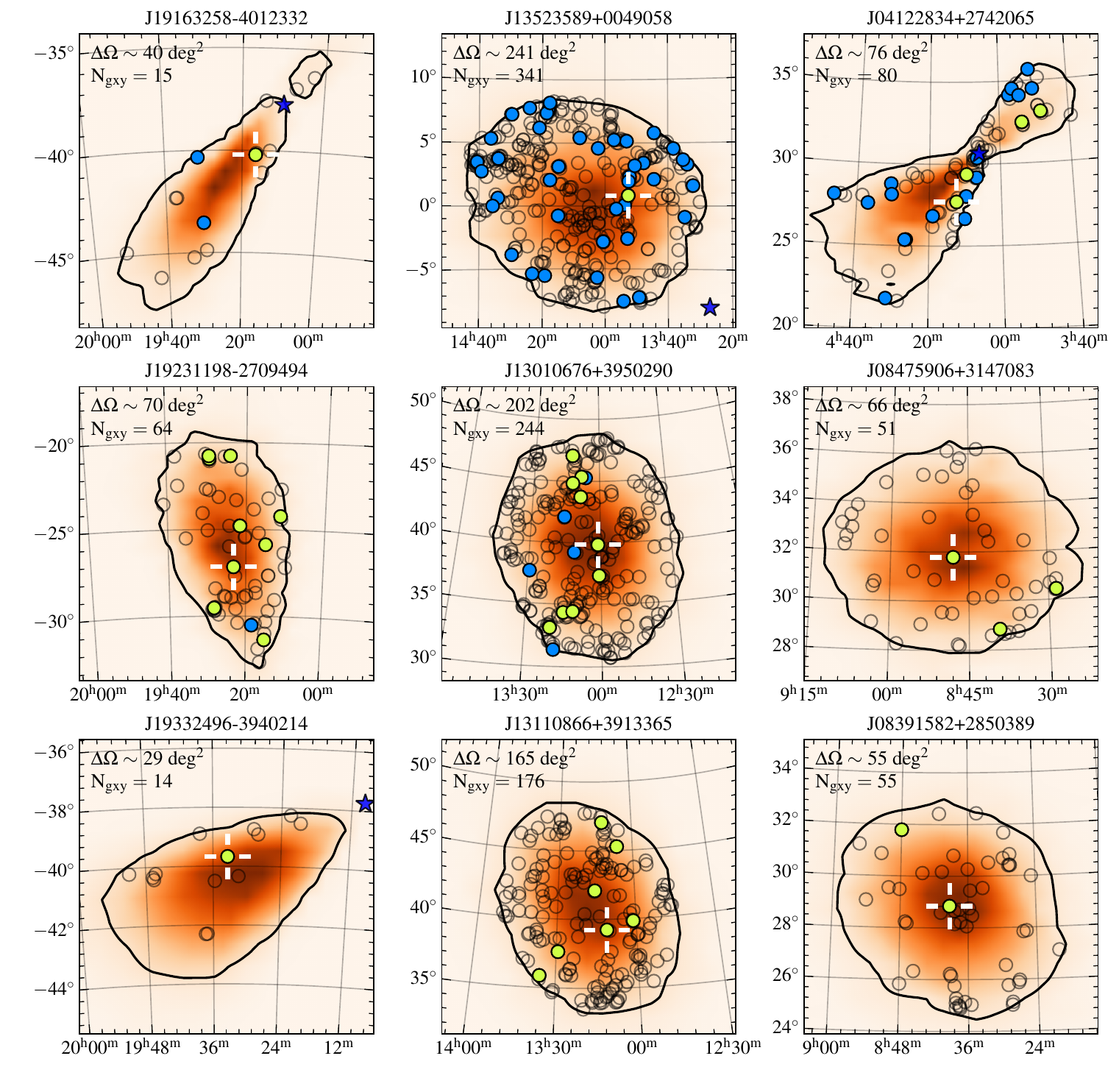}
    \caption{Zoomed-in localization areas for the SNR=15 simulation set, arranged in the same layout as in \autoref{fig:9gxy_snr15}. The probability density maps are shown in orange, with darker regions having higher probabilities, and the 90\% credible regions are outlined by the black contours. All possible host galaxies in NANOGrav's catalog that fall within the contours are represented by open circles, while galaxies passing the mass cut appear as blue circles, and galaxies passing both the mass and distance cuts appear as yellow-green circles. The true galaxy is highlighted with white crosshairs. For ease of viewing, the cuts shown here correspond to the $1\sigma$ numbers listed under the $f_{\mathrm{bulge}}=0.31$ columns in \autoref{tab:snr15_cuts}. The median localization area $\Delta\Omega$ and the number of galaxies $\mathrm{N}_{\mathrm{gxy}}$ within the area of one representative bootstrapped sky location posterior are provided in the top left corner of each panel.}
    \label{fig:9gxy_snr15_zoom}
\end{figure*}

\begin{figure}[!ht]
    \centering \includegraphics[width=0.45\textwidth]{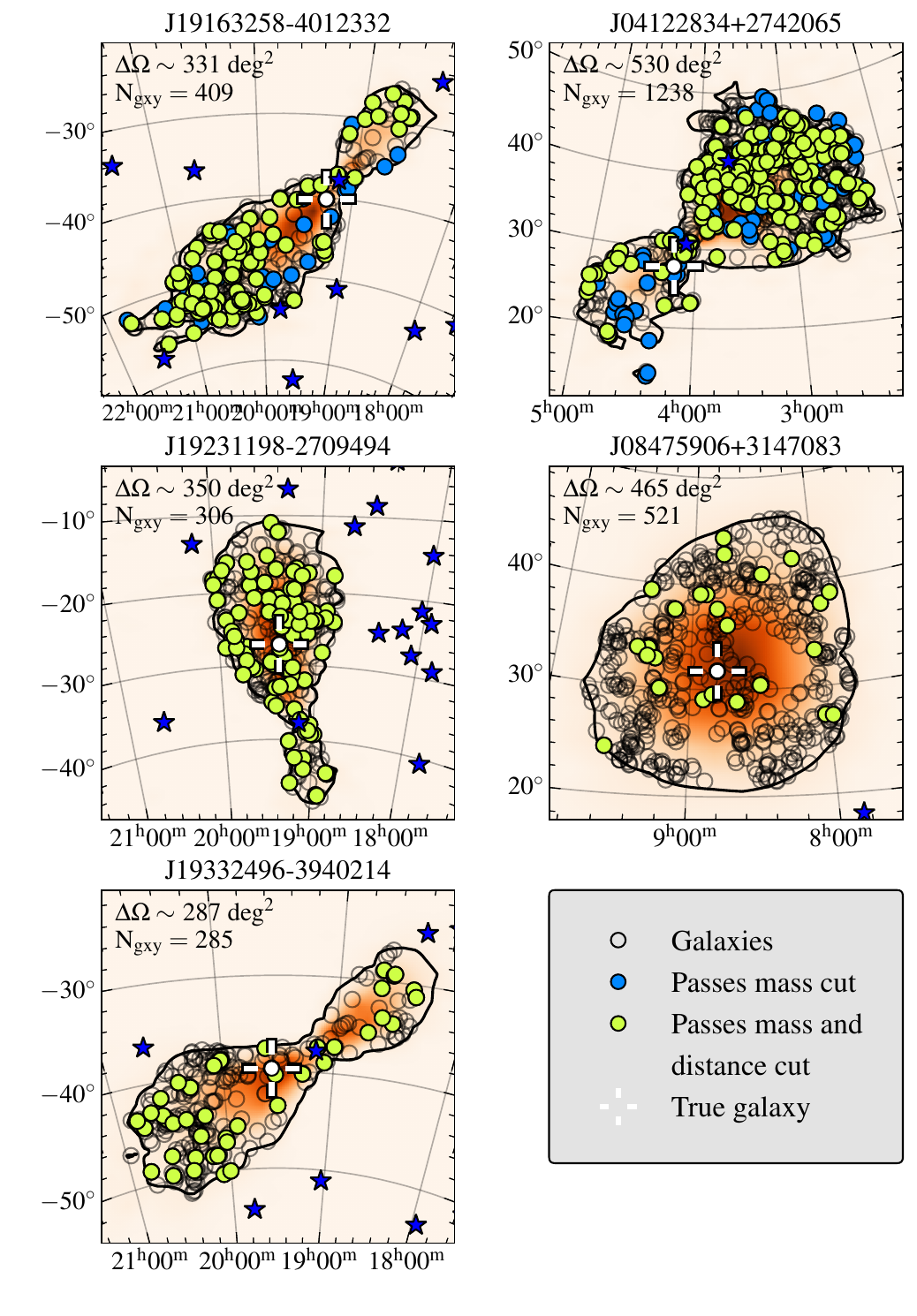}
    \caption{Zoomed-in localization areas for the SNR=8 simulation set, including only those localization areas that are well-constrained, i.e., the entire left column and first two panels in the right column of \autoref{fig:9gxy_snr8}. Each panel is formatted in the same manner as in \autoref{fig:9gxy_snr15_zoom}. For ease of viewing, these cuts correspond to the $1\sigma$ numbers listed under the $f_{\mathrm{bulge}}=0.31$ columns in \autoref{tab:snr8_cuts}.}
    \label{fig:9gxy_snr8_zoom}
\end{figure}

We investigate the number of potential hosts within the SNR=15 localization regions using the zoom-in panels shown in \autoref{fig:9gxy_snr15_zoom}. Likewise, the zoom-in panels for the five localized SNR=8 sources are shown in \autoref{fig:9gxy_snr8_zoom}. In addition to the probability density maps (orange), the 90\% credible region contours (black outlines), and any nearby pulsars (blue stars), we include all potential host galaxies within the area as circles and highlight the true galaxy into which the SMBHB was injected with white crosshairs. The median localization area $\Delta\Omega$ and the number of galaxies enclosed $\mathrm{N}_{\mathrm{gxy}}$ are listed in the top left corner of each panel. 

Both figures show that larger localization areas naturally contain a larger number of potential host galaxies. In \autoref{fig:9gxy_snr15_zoom}, the smallest area is about 29 deg$^2$ and contains 14 galaxies (bottom left), while the largest area is about 241 deg$^2$ and contains 341 galaxies (top middle). The much broader areas in \autoref{fig:9gxy_snr8_zoom} enclose upwards of $\sim$ 300 potential hosts; the smallest area here is about 287 deg$^2$ and contains 285 galaxies (again, bottom left), and largest area is about 530 deg$^2$ and contains 1238 galaxies (top right).

\begin{figure}[!ht]
    \centering
    \includegraphics[width=0.45\textwidth]{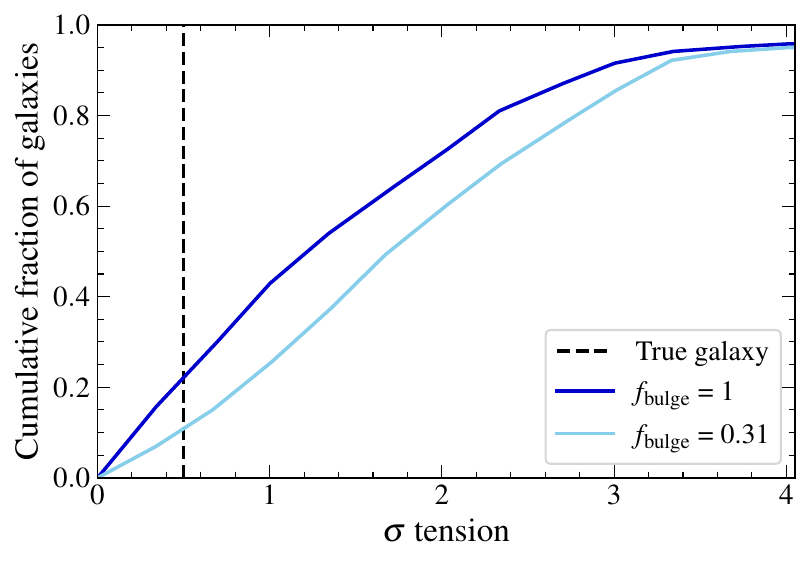}
    \caption{Cumulative distribution of SMBH mass tension values for galaxies in the SNR=8 localization region for source J04122834+2742065 (shown in the top right panel of \autoref{fig:9gxy_snr8_zoom}). Because a small percentage of galaxies have infinite sigma tension with the GW posterior, the distributions do not reach a cumulative fraction of 1. The solid dark blue and light blue curves correspond to all unknown-type galaxies having SMBH masses estimated with $f_{\mathrm{bulge}}=1$ and $f_{\mathrm{bulge}}=0.31$, respectively. The tension value for the true galaxy is indicated by the vertical dashed line.}
    \label{fig:tens_CDF}
\end{figure}

The number of potential hosts is pared down by implementing the full pipeline outlined in Section \ref{sec:cuts}. We take all galaxies enclosed within the 90\% credible area and compute the tension between each galaxy's SMBH mass estimate and the total binary mass posterior. The cumulative distribution of tensions for one example SNR=8 source, J04122834+2742065, is presented in \autoref{fig:tens_CDF}. Of the 1238 galaxies within this source's localization region, 5$-$6\% are in ``infinite" sigma tension with the GW posterior. These galaxies have SMBH masses $\leq 10^{7.6}$ $\mathrm{M}_\odot$, while the SMBH mass of the injected source is $10^{9.31}$ $\mathrm{M}_\odot$; thus, infinite tension arises when the two distributions overlap very little, or not at all. To account for such galaxies, the distributions in \autoref{fig:tens_CDF} only extend up to the cumulative fraction of galaxies with finite sigma tension.

Because nearly half of the galaxies in the catalog have unknown morphological types, we include two distributions, one for which all unknown-type galaxies have SMBH masses estimated with $f_{\mathrm{bulge}}=1$ (solid dark blue curve) and the other with $f_{\mathrm{bulge}}=0.31$ (solid light blue curve), corresponding to elliptical and Sa galaxies, respectively. While we typically do not expect to observe binaries in spiral galaxies due to the galaxy merger's destruction of the disk, these two scenarios act as a way to quantify the uncertainty on the bulge fraction. The distribution using $f_{\mathrm{bulge}}=0.31$ is generally in higher tension with the GW posterior compared to that with $f_{\mathrm{bulge}}=1$, as smaller bulge masses will naturally yield smaller SMBH masses according to the $M_{\mathrm{BH}}-M_{\mathrm{bulge}}$ relation. Consequently, the $f_{\mathrm{bulge}}=0.31$ curve represents an optimistic scenario for the identification of the host galaxy, while the $f_{\mathrm{bulge}}=1$ curve is conservative, in that placing a threshold at any tension value will result in fewer remaining galaxies when using $f_{\mathrm{bulge}}=0.31$ over $f_{\mathrm{bulge}}=1$.

We also mark the tension between the true galaxy's SMBH mass and the total binary mass posterior with the vertical dashed line. Here, and across all of our sources in both sets of simulations, this tension value is consistently $< 1\sigma$. As example cuts on the tens to hundreds of galaxies in these localization areas, we therefore implement a tension threshold of $1\sigma$, as well as a more conservative threshold of $2\sigma$, and discard any galaxies with tensions above these values. Following this step, we include a final cut on the luminosity distance and remove galaxies farther than the $95^\mathrm{th}$ percentile of the distance posterior.

In \autoref{fig:9gxy_snr15_zoom} and \autoref{fig:9gxy_snr8_zoom}, we color the galaxies within each sky area to distinguish between those that simply lie within the localization area (open circles), versus those that pass the mass cut (blue circles), and those that pass both the mass and distance cuts (yellow-green circles). \autoref{tab:snr15_cuts} and \autoref{tab:snr8_cuts} contain a complete summary of the number of galaxies remaining after each successive cut when using both the $f_{\mathrm{bulge}}=1$ and $f_{\mathrm{bulge}}=0.31$ SMBH mass estimates. When reducing the number of potential hosts, the SMBH mass cuts are evidently much more effective than the distance cuts. While the distance cuts help to cut down on hosts mainly for sources closer than 100 Mpc, sources between 100$-$300~Mpc and farther than 300 Mpc benefit from distance cuts only slightly, if at all. This difference in efficacy of the two cuts is reasonable, as the uncertainties on the total binary mass obtained from the CW search are smaller (ranging from 0.13$-$0.41~dex, or fractional uncertainties between 1$-$5\%), and therefore more informative, than the uncertainties on the distance posteriors (0.2$-$0.7~dex, or fractional uncertainties as high as 34\%). Changing the order of the cuts also does not make a difference, either in their efficacy or in the final number of galaxies remaining.

Since the SNR=15 localization regions are smaller and enclose fewer galaxies to begin with, there are sensibly far fewer galaxies left after cuts as compared to the SNR=8 examples. At worst, 22 galaxies pass the final cut, and at best, only one galaxy -- the true galaxy into which the CW signal was injected -- remains when using a $1\sigma$ SMBH mass cut. With a $2\sigma$ SMBH mass cut, these numbers increase to 81 and 2 for the worst and best cases, respectively. While these results are promising, the SNR=15 scenario is optimistic, as the first CW signal detected by PTAs will not likely boast such a high SNR. For example, \cite{2015MNRAS.451.2417R}, \cite{2018MNRAS.477..964K}, and \cite{2022ApJ...941..119B} predict SNRs $\lesssim$ 5 for simulated datasets of similar observing timespans. In this case, a signal detection may be possible, but it would not guarantee well-constrained binary parameters. However, these SNR estimates may be somewhat conservative, as the aforementioned studies involve PTAs simulated with 42$-$58\% of the number of pulsars used in this work. Regardless, our SNR=8 sources represent a more realistic scenario. If the source can be localized at all, i.e., it has enough pulsars nearby to help constrain the sky location, we may contend with as many as $\sim$~400 ($\sim$~730) potential galaxy hosts even after implementing $1\sigma$ ($2\sigma$) cuts. Such numbers, and even something like the best case scenario of $\sim$~30 ($\sim$~130) galaxies, will require additional criteria to further cut down on hosts. We discuss future plans in this direction in Section \ref{sec:discuss_cuts}.

\begin{deluxetable*}{ccccccc}
\setlength{\tabcolsep}{1em}
\tablecaption{Summary of the host galaxy cut procedure for the SNR=15 sources. The columns show the median localization areas (with uncertainties indicating the $16^\mathrm{th}$ and $84^\mathrm{th}$ percentiles), the number of galaxies $\mathrm{N}_{\mathrm{gxy}}$ within the localization region of a representative bootstrapped sky location posterior, the number remaining after a 1$\sigma$ (2$\sigma$) cut on the SMBH mass, and the number remaining after a successive cut on the luminosity distance. The cuts are done using both the $f_{\mathrm{bulge}}=1$ and $f_{\mathrm{bulge}}=0.31$ SMBH mass estimates from the galaxy catalog, which roughly represent conservative and optimistic scenarios, respectively. In each case, the true galaxy passes all cuts; for sources where the remaining number of galaxies is 1, this is indeed the correctly identified host galaxy.}\label{tab:snr15_cuts}
\tablecolumns{7}
\tablehead{
\multirow{2}{*}{Galaxy ID} & \multirow{2}{*}{$\Delta\Omega$ (deg$^2$)} & \multirow{2}{*}{$\mathrm{N}_{\mathrm{gxy}}$} & \multicolumn{2}{c}{Mass cut} & \multicolumn{2}{c}{Mass \texttt{+} distance cut} \\
\cline{4-7}
\colhead{} & \colhead{} & \colhead{} & \colhead{$f_{\mathrm{bulge}}=1$} & \colhead{$f_{\mathrm{bulge}}=0.31$} & \colhead{$f_{\mathrm{bulge}}=1$} & \colhead{$f_{\mathrm{bulge}}=0.31$}
}
\startdata
J19163258$-$4012332&$39.9_{-0.4}^{+0.4}$&15&4 (6)&3 (5)&1 (3)&1 (3)\\
J19231198$-$2709494&$70.1_{-0.4}^{+0.5}$&64&23 (42)&10 (25)&22 (41)&9 (24)\\
J19332496$-$3940214&$29.0_{-0.1}^{+0.1}$&14&2 (4)&1 2)&2 (3)&1 (2)\\
J13523589+0049058&$241_{-2}^{+2}$&341&72 (164)&43 (109)&1 (4)&1 (4)\\
J13010676+3950290&$202_{-2}^{+1}$&244&37 (117)&14 (74)&9 (81)&9 (50)\\
J13110866+3913365&$165_{-1}^{+1}$&176&17(65)&7 (38)&17 (65)&7 (38)\\
J04122834+2742065&$76.5_{-0.8}^{+1.1}$&80&28 (44)&21 (38)&4 (7)&4 (7)\\
J08475906+3147083&$65.9_{-0.4}^{+0.5}$&51&8 (21)&3 (11)&5 (18)&3 (8)\\
J08391582+2850389&$54.6_{-0.5}^{+0.5}$&55&7 (17)&2 (9)&6 (15)&2 (7)\\
\enddata
\end{deluxetable*}

\begin{deluxetable*}{ccccccc}
\setlength{\tabcolsep}{1em}
\tablecaption{Summary of the host galaxy cut procedure for the SNR=8 sources. Only the five constrained sources in this set are shown. Again, in each case, the true galaxy passes all cuts.\label{tab:snr8_cuts}}
\tablecolumns{7}
\tablehead{
\multirow{2}{*}{Galaxy ID} & \multirow{2}{*}{$\Delta\Omega$ (deg$^2$)} & \multirow{2}{*}{$\mathrm{N}{\mathrm{gxy}}$} & \multicolumn{2}{c}{Mass cut} & \multicolumn{2}{c}{Mass \texttt{+} distance cut} \\
\cline{4-7}
\colhead{} & \colhead{} & \colhead{} & \colhead{$f_{\mathrm{bulge}}=1$} & \colhead{$f_{\mathrm{bulge}}=0.31$} & \colhead{$f_{\mathrm{bulge}}=1$} & \colhead{$f_{\mathrm{bulge}}=0.31$}
}
\startdata
J19163258$-$4012332&$331_{-4}^{+4}$&409&203 (331)&164 (289)&167 (292)&123 (250)\\
J19231198$-$2709494&$350_{-7}^{+6}$&306&173 (264)&96 (225)&173 (264)&96 (225)\\
J19332496$-$3940214&$287_{-4}^{+4}$&285&82 (181)&43 (147)&82 (181)&43 (147)\\
J04122834+2742065&$530_{-8}^{+7}$&1238&531 (890)&312 (738)&397 (733)&214 (581)\\
J08475906+3147083&$465_{-5}^{+6}$&521&51 (194)&27 (126)&51 (194)&27 (126)\\
\enddata
\end{deluxetable*}

\section{Discussion} \label{sec:discuss}

Our results necessarily rely on a number of assumptions and choices, which we discuss in further detail here, as well as comparisons to previous studies and future plans to expand on this work.

\subsection{Comparisons to Previous Studies} \label{sec:discuss_comp}

In Section \ref{sec:results_areafac} we saw that the size of the localization area is influenced not only by the SNR of the signal, but also by the proximity of nearby pulsars to the GW source. \citet[][hereafter \citetalias{2019MNRAS.485..248G}]{2019MNRAS.485..248G} have identified and explored a similar trend, which we now briefly compare to our findings. Figure 3 from \citetalias{2019MNRAS.485..248G} is analogous to our \autoref{fig:area_vs_snr} in that both present localization areas recovered for a range of injected SNRs and for three sources located in the different ``sensitivity regions" of the sky. Across all three sources, we find that our sky areas are generally smaller than those in  \citetalias{2019MNRAS.485..248G}, which could be due to a few factors. 

The improvements in localization capability may stem in part from the updated sensitivity of our simulated PTA, as our IPTA-DR3 version contains more than double the number of pulsars and roughly double the observing timespan of the IPTA-DR1 version in \citetalias{2019MNRAS.485..248G}. We also make use of the pulsars’ real observation baselines and noise properties, while \citetalias{2019MNRAS.485..248G} adjust these factors to achieve their desired SNR. Thus, although the signal SNRs are similar between the two studies, the different PTA configurations may lead to differences in the signal localization.

In addition to this, we may see improvements in the localization areas due to the different GW analysis methods employed. \citetalias{2019MNRAS.485..248G} use a null-stream analysis involving a three-dimensional likelihood function of the amplitude and sky location. This analysis does not implement the full CW signal with both the earth term and pulsar term components (as in \autoref{eq:s_t}) but is rather an earth-term-only analysis \citep{2018MNRAS.477.5447G}. However, accounting for the pulsar term is essential in all-sky CW searches, not only in determining the sky location more accurately where earth-term-only searches may be biased \citep{2016MNRAS.461.1317Z}, but also in improving constraints on the sky location \citep{2011MNRAS.414.3251L,2010arXiv1008.1782C}. Consequently, our full signal analyses may yield smaller localization areas.

In spite of these differences, the dependence of GW source localization on the proximity of nearby pulsars is clear. \citetalias{2019MNRAS.485..248G} and \cite{2010PhRvD..81j4008S} note that CW sources are best localized in regions where there are many pulsars as well as in antipodal regions; on the other hand, sources in the middle region of the sky suffer from poor localization as they sit orthogonal to most pulsars. In our study and in \citetalias{2019MNRAS.485..248G}, we see that the anisotropic distribution of pulsars on the sky indeed causes sources in the ``middle sensitivity" region to be the worst localized. Further, \citetalias{2019MNRAS.485..248G} quantify the localization area across the sky for sources of fixed SNR (see Figure 4), while \autoref{fig:area_vs_psrangsep} of this work relates the localization area to the angular separation between the source and the nearest pulsars.

\subsection{Galaxy catalog completeness}

For our most distant sources lying at $d_L >$ 300 Mpc, the SNR=15 GW posteriors lie just within the distribution of distances in the galaxy catalog, while the SNR=8 posteriors extend well beyond the catalog distances. Thus, although the galaxy catalog used throughout this work reaches out to 500$-$700~Mpc for the most massive galaxies targeted by PTAs, the catalog's completeness still poses an issue. By default, the true host galaxy of the GW source is always included in our catalog, but this assumption may not hold true for a real GW detection. Multiple studies predict that the first individual SMBHBs seen by PTAs will lie at Gpc distances \citep{2015MNRAS.451.2417R,2018MNRAS.477..964K,2022ApJ...941..119B}, making it a real possibility that we will detect a GW source beyond the limits of our current galaxy catalog. However, these studies also predict that the most detectable sources will have relatively high chirp masses (and therefore high galaxy stellar masses), and galaxy catalogs are typically the most complete for such galaxies.

As is typical of many all-sky surveys, our catalog also avoids the galactic plane. At the same time, the galactic plane is where the majority of our pulsars are observed. These two facts pose another obstacle: the region of the sky containing the most pulsars makes it the most sensitive to individual binaries, but this region is also most likely incomplete in our galaxy catalog. Although we may find a well-localized GW source in the high sensitivity region of the sky, the source's true host may be absent from the catalog.

Both of these gaps -- the depth as well as the sky coverage -- could produce a scenario in which our host identification pipeline puts forth either a false candidate or no viable candidates at all. In the future, if we want to maximize our chances of identifying the correct host, it will be vital not only to expand current galaxy catalogs to larger volumes, but also to fill in areas of the sky obscured by the galactic plane. While some all-sky surveys do cover the galactic plane and could be used to fill these gaps, these surveys often lack quantities like the SMBH mass; therefore, if such surveys are added to our catalog, the number of potential hosts may increase, but we may not be able to implement effective cuts on them. 

\subsection{SMBH mass uncertainties}

We showed in Section \ref{sec:results_cuts} that we see a significant decrease in the number of potential hosts when we introduce a filter on the SMBH mass. Even so, the SMBH masses in our galaxy catalog have varied levels of uncertainty, as the methods employed to calculate them are not uniform across all galaxies. 95\% of the galaxies in the catalog have SMBH masses estimated using the $M_{\mathrm{BH}} - M_{\mathrm{bulge}}$ relation. In general, obtaining SMBH mass estimates from such global scaling relations is more feasible and less expensive to do than dynamical measurements, but these estimates have intrinsic scatter, resulting in the largest uncertainties. A larger uncertainty on the SMBH mass will translate to greater overlap with the GW posterior and, consequently, better agreement between the two distributions. This would ultimately yield a larger number of potential hosts remaining after cuts.  While more precise SMBH mass measurements could help to discard additional galaxies, attaining more direct measurements is not feasible for such a large sample of galaxies. A separate estimate of SMBH mass can be obtained from the $M_{\mathrm{BH}} - \sigma$ relation, but spectroscopy would be required to measure galaxy velocity dispersions.

Another source of SMBH mass uncertainty is that 45\% of galaxies with $M_{\mathrm{BH}} - M_{\mathrm{bulge}}$ estimates have unknown morphological types. The morphological type informs the fraction of stellar mass in the galaxy's bulge $f_{\mathrm{bulge}}$, which is then used to calculate the SMBH mass. For unknown-type galaxies, this consequently introduces some additional uncertainty on the SMBH mass. Our galaxy catalog attempts to quantify this uncertainty by assuming two possibilities for unknown type galaxies -- one SMBH mass estimate corresponding to an elliptical galaxy and one SMBH mass estimate corresponding to an Sa-type spiral galaxy -- both of which we incorporate into our cut procedure.

By assuming that all unknown-type galaxies are elliptical, we present the most conservative case; all other morphology assumptions would result in smaller SMBH masses, have higher tension with the GW posterior, and ultimately yield fewer galaxies remaining after cuts. Therefore, if we can obtain a galaxy catalog where the morphological types are known (or can be determined), the real number of hosts remaining after cuts will lie somewhere between the two distributions in \autoref{fig:tens_CDF}. Otherwise, if the morphological types are not known from the catalog, they could be acquired with follow-up observations for a small selection of plausible hosts. Overall, though, the uncertainty in galaxy morphology is likely to be small in this work as only SMBHs above $\sim 10^9$~${\rm M}_\odot$ are considered. The host galaxies of these SMBHs have stellar masses above $\sim 10^{11.5}$~${\rm M}_\odot$, a mass range in which at least $\sim 80$\% of local galaxies are quiescent elliptical galaxies \citep[e.g.,][]{2013ApJ...767...50M}.

\subsection{Cut criteria} \label{sec:discuss_cuts}

In Section \ref{sec:results_cuts} we included uncertainties on the sky areas recovered for well-localized sources, which we estimated using bootstrapped posterior distributions. The random resampling involved in bootstrapping causes each new set of posterior samples to differ slightly from the next; the uncertainty on the size of the area therefore suggests that there exists some uncertainty on the number of galaxies as well. However, we chose not to include uncertainties on the number of potential hosts within these regions but rather provided these numbers for one representative bootstrapped sky location posterior for each of our sources. 

This choice was made for a couple reasons. First, the uncertainty on the number of potential hosts is relatively small across all of our injections. To place uncertainties on these numbers, we calculated the number of galaxies enclosed within the localization area of each bootstrapped sky location posterior. From the distribution of these numbers, we then took the $16^\mathrm{th}$ and $84^\mathrm{th}$ percentiles as the lower and upper uncertainties, finding that the fractional uncertainty on the number of hosts is on the order of 2$-$3\% for both simulation sets. Second, the gaps in our galaxy catalog pose an issue here. Since the catalog used throughout this work is incomplete, the numbers we quote can be considered as lower bounds on the total number of potential hosts.

Aside from this, our aim in this work is not to establish the best method of cutting down on potential host galaxies, but rather to simulate end-to-end the GW recovery and host identification prospects using a realistic pipeline and galaxy catalog. In reality, there are many different cut approaches one could take. It would be interesting to explore different cut methods and compare their efficacy in both reducing the number of hosts and identifying the true host. For example, we choose to discard galaxies lying farther than the $95^\mathrm{th}$ percentile of the luminosity distance posterior, but this cut is only effective for nearby sources. For distant sources, an equal-tailed distance cut may be more effective in discarding hosts.

Our cuts also do not assume any covariance between the GW parameters. Most significantly, measurement covariance exists between the chirp mass and luminosity distance, particularly for binaries of low mass or low GW frequency, as these parameters are related to the GW strain. It may therefore be more appropriate to perform a simultaneous, two-dimensional cut on the total binary mass and luminosity distance together. Though we tried such cuts using the \texttt{tensiometer} package, our results showed variable performance in comparison to the one-dimensional cuts done throughout this work, and we leave this approach as a subject for future study. Similarly, our cuts assume that there is no covariance between the sky location and distance posteriors, making our localization volume effectively cylindrical, whereas a more sophisticated cut could incorporate the true three-dimensional localization volume. Finally, given that the SNR=8 case concluded with as many as $\sim$~400 ($\sim$~730) galaxies remaining after stringent (conservative) cuts, it may be beneficial to add further cut criteria, potentially involving information from EM observations \citep[e.g.,][]{2024ApJ...961...34B, 2024arXiv240714061B, 2024arXiv240701659C, 2024MNRAS.529.4295S}.

\section{Conclusions} \label{sec:concl}

In this paper, we carried out a ``dress rehearsal" pipeline to quantify the identification prospects of SMBHB host galaxies, combining simulated IPTA-DR3-type datasets with NANOGrav's catalog of massive galaxies in the local universe. We selected nine galaxies to inject with a putative SMBHB and created two sets of simulated datasets, one with CW signals of SNR=8 and the other with SNR=15. After recovering the signal and estimating the binary parameters, we used NANOGrav's galaxy catalog to determine the approximate number of galaxies contained within each recovered localization area. Finally, the number of plausible hosts in the 90\% credible region was reduced by setting thresholds on the SMBH mass and luminosity distance based on our recovered GW posteriors.

For the SNR=8 sources that are well-constrained, the localization areas range from $\sim$ 287 deg$^2$ to 530 deg$^2$, containing about 285 to 1238 potential hosts. After implementing both stringent and conservative cuts, we are left with anywhere from about 27 to 397 hosts, or 126 to 733 hosts, respectively. The better-constrained SNR=15 sources naturally have smaller areas that contain fewer galaxies; within areas ranging from $\sim$ 29 deg$^2$ to 241 deg$^2$, the number of potential hosts varies from about 14 to 341. This number is reduced to about 22 hosts at worst and 1 true host at best after stringent cuts, or about 81 hosts at worst and 2 hosts at best after conservative cuts. While the SNR=15 examples are promising for the era in which we have already detected a CW signal and are working to better resolve it, the SNR=8 case is more realistic for the first CWs that will be detected and localized by PTAs. Further, a threshold SNR~$\sim$~8 is necessary for the data to be informative enough in estimating the sky location. To successfully narrow down the list of host candidates at such low SNRs, though, further galaxy criteria, reduction methods, or dedicated EM follow-up observations will be crucial. 

In addition to studying these sources, we conducted a number of case studies to more deeply explore the parameters governing the size of the localization area. While the localization area is primarily determined by the SNR of the CW signal, it does not act alone; we find that two other variables also influence the size of the area: the proximity of the source to nearby pulsars on the sky and the source's chirp mass.

The impact of nearby pulsars is particularly evident when comparing localization areas for sources in different ``sensitivity" regions of the sky. We define the sensitivity as the ability to detect a signal, separate from determining the signal's location. For a fixed set of binary parameters injected across the sky, the high sensitivity region sees the highest expected SNR from a GW signal. Because the majority of pulsars observed by the IPTA lie in this region, GW signals found here are typically close to many nearby pulsars and naturally tend to have the smallest localization areas. However, sources in the antipodal low sensitivity region, where the expected SNR would be lowest, have the next best localization, outperforming sources lying in a region with middling sensitivity.

This is a direct result of the fact that the mid sensitivity region of the sky is distinctly lacking in pulsars. Although the low sensitivity region contains few pulsars, GWs coming from this region are generally closer to pulsars on the sky than those in the mid sensitivity region. It is therefore important to distinguish between sensitivity and localization: when a signal is detected in the low sensitivity region, we may have some confidence in determining its location, but detecting the signal in the first place is more challenging in this region than in other parts of the sky. The reverse is true for the mid sensitivity region: it may be easier to detect a signal here, but the localization capability of PTAs in this region is weakest. We show that this dependence on the closest pulsars can be distilled into a positive relationship between their angular separation from the GW source and the recovered localization area. Similar results have previously been shown in \cite{2019MNRAS.485..248G}, aside from some minor differences in PTA configurations and GW analyses.

We also show that the chirp mass of the binary plays a role in signal localization. This effect is apparent only when the chirp mass is high enough to produce substantial evolution between the GW frequency of the earth term and that of the pulsar term. For injections of SNR=20 and an earth term frequency of $f_{\mathrm{GW}} = 20$~nHz, we determined that the chirp mass contributes to the localization around $\log_{10}(\mathcal{M}/\mathrm{M}_\odot) \gtrsim 9.3$, at which point the sky area decreases with increasing chirp mass. However, this ``threshold" value is likely different for signals of different earth term frequencies. SMBHBs producing GWs at frequencies lower than $20$~nHz (or at wider orbital separations) will evolve more slowly, spending more time in a given frequency bin. In order for such a binary to evolve enough that the earth and pulsar term frequencies are significantly different and provide more information about the source's location, the binary's chirp mass would need to be higher than $\log_{10}(\mathcal{M}/\mathrm{M}_\odot) \gtrsim 9.3$. Conversely, binaries with frequencies higher than $20$~nHz could start to see additional position constraints at lower chirp masses. We plan to further explore the relative importance of these parameters and tease out more general relationships connecting them to the size of the localization area in a separate study.

Finally, our work aims to begin preparations for the era in which PTAs will pick up on individual SMBHBs. As this is only the beginning, there are a number of ways in which our host galaxy identification pipeline can be improved. Starting with the host galaxies themselves, our existing pool of possible hosts provided by NANOGrav’s galaxy catalog may not be entirely complete. Galaxy catalogs used for SMBHB host identification will need upgrades in both the sky coverage and depth, as well as more complete morphological classification of galaxies; otherwise, the hosts of GW sources sitting around the galactic plane or beyond $\sim$~500 Mpc will not be identified efficiently. While follow-up observations for these sources can be triggered, such observations are time-consuming, and our objective is rather to prepare for host identification as much as possible ahead of a GW detection.

Aside from the issue of completeness, the criteria used to cut down on potential hosts can be improved as well. Cuts on the SMBH mass are particularly effective in reducing the number of hosts, but the majority of SMBH mass estimates in the galaxy catalog have fairly large uncertainties, which come from the $M_{\mathrm{BH}} - M_{\mathrm{bulge}}$ scaling relation. We could filter out more hosts if the SMBH masses can be determined more precisely, either as part of a catalog or through EM follow-up observations on galaxies remaining after cuts. SMBH masses derived from the $M_{\mathrm{BH}} - \sigma$ relation, for example, would not require knowledge of the morphological type, but spectroscopic data would be required to determine galaxy velocity dispersions. In the future, we plan to investigate options for additional cut criteria, and as a final step in our pipeline, we aim to implement a ranking scheme for host candidates remaining after cuts.

\begin{acknowledgments}
We thank our colleagues in NANOGrav for useful feedback on the manuscript. We particularly thank Bence Bécsy and Neil Cornish for fruitful discussions, and Nihan Pol for assisting with the creation of our simulated PTA datasets. PP acknowledges support from NASA FINESST grant number 80NSSC23K1442. SRT and MC acknowledge support from NSF AST-2007993. SRT acknowledges support from an NSF CAREER \#2146016. SRT and CPM acknowledge support from NSF AST-2307719. MC acknowledges support by the European Union (ERC, MMMonsters, 101117624). The authors are members of the NANOGrav collaboration, which receives support from NSF Physics Frontiers Center award number 1430284 and 2020265. This work was conducted in part using the resources of the Advanced Computing Center for Research and Education (ACCRE) at Vanderbilt University, Nashville, TN.
\end{acknowledgments}

\software{\texttt{enterprise} \citep{2019ascl.soft12015E}, \texttt{HEALPix} \citep{2005ApJ...622..759G}, \texttt{healpy} \citep{Zonca2019}, \texttt{Jupyter} \citep{soton403913}, \texttt{libstempo} \citep{2020ascl.soft02017V}, \texttt{ligo.skymap} \citep{2016PhRvD..93b4013S}, \texttt{matplotlib} \citep{2007CSE.....9...90H}, \texttt{numpy} \citep{2020Natur.585..357H}, \texttt{PTMCMCSampler} \citep{justin_ellis_2017_1037579}, \texttt{QuickCW} \citep{2022PhRvD.105l2003B}, \texttt{tensiometer} \citep{2021PhRvD.104d3504R}}

\bibliography{ref}{}
\bibliographystyle{aasjournal}

\end{document}